\documentclass[fleqn,usenatbib]{mnras}

\def\degr{\hbox{$^\circ$}}
\def\arcmin{\hbox{$^\prime$}}
\def\arcsec{\hbox{$^{\prime\prime}$}}

\def\micron{\hbox{$\mu$m}}
\def\n2hp{\hbox{N$_2$H$^+$(1-0)}}

\usepackage{natbib}
\usepackage{rotating}
\usepackage{graphicx}
\usepackage{capt-of}
\usepackage{tablefootnote}
\usepackage{booktabs,caption}
\usepackage[flushleft]{threeparttable}
\usepackage{amsmath}
\usepackage{txfonts}

\usepackage[T1]{fontenc}

\DeclareRobustCommand{\VAN}[3]{#2}
\let\VANthebibliography\thebibliography
\def\thebibliography{\DeclareRobustCommand{\VAN}[3]{##3}\VANthebibliography}

\usepackage{graphicx}	
\usepackage{amsmath}	
\usepackage{amssymb}	

\title[Star cluster progenitors are dynamically decoupled]{Star cluster progenitors are dynamically decoupled from their parent molecular clouds}

\author[N. Peretto]{
Nicolas Peretto$^{1}$\thanks{E-mail: nicolas.peretto@astro.cf.ac.uk}
Andrew J. Rigby $^{2}$
Fabien Louvet $^{3}$
Gary A. Fuller$^{4,5}$
Alessio Traficante$^{6}$
Mathilde Gaudel$^7$
\\
$^1$ Cardiff Hub for Astrophysics Research \& Technology, School of Physics \& Astronomy, Cardiff University, Queens Buildings, The parade, Cardiff CF24 3AA, UK\\
 $^2$ School of Physics and Astronomy, University of Leeds, Leeds LS2 9JT, UK\\
$^3$ Univ. Grenoble Alpes, CNRS, IPAG, 38000 Grenoble, France\\
$^4$Jodrell Bank Center for Astrophysics, Department of Physics \& Astronomy, University of Manchester, Oxford Road,Manchester, M13 9PL, UK\\
$^5$ I. Physikalisches Institut, University of Cologne, Z\"ulpicher Str. 77, 50937 K\"oln, Germany\\
$^6$ IAPS-INAF, Via Fosso del Cavaliere, 100, I-00133 Rome, Italy\\
$^7$ LERMA, Observatoire de Paris, PSL Research University, CNRS, Sorbonne Universit\'e, 75014, Paris, France
}

\date{Accepted 5$^{\rm{th}}$ of August 2023. Received  17$^{\rm{th}}$ of March 2023; in original form ZZZ}

\pubyear{2015}

\begin{document}
\label{firstpage}
\pagerange{\pageref{firstpage}--\pageref{lastpage}}
\maketitle

\begin{abstract}
The formation of stellar clusters dictates the pace at which galaxies evolve, and solving the question of their formation will undoubtedly lead to a better understanding of the Universe as a whole. While it is well known that star clusters form within parsec-scale over-densities of interstellar molecular gas called clumps, it is, however, unclear whether these clumps represent the high-density tip of a continuous gaseous flow that gradually leads towards the formation of stars, or a transition within the gas physical properties. Here, we present a unique analysis of a sample of 27 infrared dark clouds embedded within 24 individual molecular clouds that combine a large set of observations, allowing us to compute the mass and velocity dispersion profiles of each, from the scale of tens of parsecs down to the scale of tenths of a parsec. These profiles reveal that the vast majority of the clouds, if not all,  are consistent with being self-gravitating on all scales, and that the clumps, on parsec-scale, are often dynamically decoupled from their surrounding molecular clouds, exhibiting steeper density profiles ($\rho\propto r^{-2}$) and flat velocity dispersion profiles ($\sigma\propto r^0$), clearly departing from Larson's relations. These findings suggest that the formation of star clusters correspond to a transition regime within the properties of the self-gravitating molecular gas. We propose that this transition regime is one that corresponds to the gravitational collapse of parsec-scale clumps within otherwise stable molecular clouds.
\end{abstract}




\section{Introduction}

Only a few years after the first detection of interstellar carbon monoxide, \cite{zuckerman1974} showed that if all the gas within dense interstellar clouds were to be freely collapsing as a
result of their self-gravity then the star formation rate in the Milky Way should be $\sim 300$~M$_{\odot}$/yr, two orders of magnitude
larger than what it actually is  \citep[$\sim 2$~M$_{\odot}$/yr - e.g.][]{robitaille2010}. In other words, molecular clouds  convert only $\sim1\%$ of their mass into stars every cloud free-fall time, making star formation a very inefficient process  \citep[e.g.][]{krumholz2007}. Despite five decades of star formation research, the physics behind this fundamental
property of molecular clouds remain to be fully understood. Over the years, a number of competing theories have been developed to explain the
low star formation efficiency of molecular clouds. The main differences between those models reside in both the fraction of the volume/mass of any molecular cloud that undergoes gravitational collapse, along with the dynamical state of the gas that  does not. In one family of models, supersonic turbulence is the one mechanism responsible for defining the mass reservoirs accessible to individual protostars and, as a result, for setting the stellar initial mass function \citep[e.g.][]{padoan1997,padoan2020,krumholz2005,hennebelle2008,hopkins2012}. In those models, the low star formation efficiency is explained by the fact that those mass reservoirs represent only a couple of percents of the molecular gas mass, the rest of the gas is either unbound or in quasi-static equilibrium and therefore does not directly participate to star formation. On the other hand, other models predict that the hierarchical gravitational collapse of molecular clouds is what drive their evolution \citep[e.g.][]{hartmann2007,ballesteros2011,vazquez-semadeni2017,vazquez-semadeni2019} and that massive star formation benefits from the favourable conditions generated by the global collapse of  dense clumps \citep[e.g.][]{bonnell2006,peretto2007,smith2009}. In those models, what limits the efficiency of star formation is stellar feedback from young low- and high-mass stars, by stabilising or dispersing most of the molecular cloud's mass \citep[e.g][]{nakamura2007,wang2010,dale2012,kim2018,offner2018,grudic2022}. The controversy around which of these two very different scenarios of star formation describes reality best fuels the majority of the star formation research for the past 20 years or so.

\begin{figure*}
   \centering
   \includegraphics[width=17.5cm]{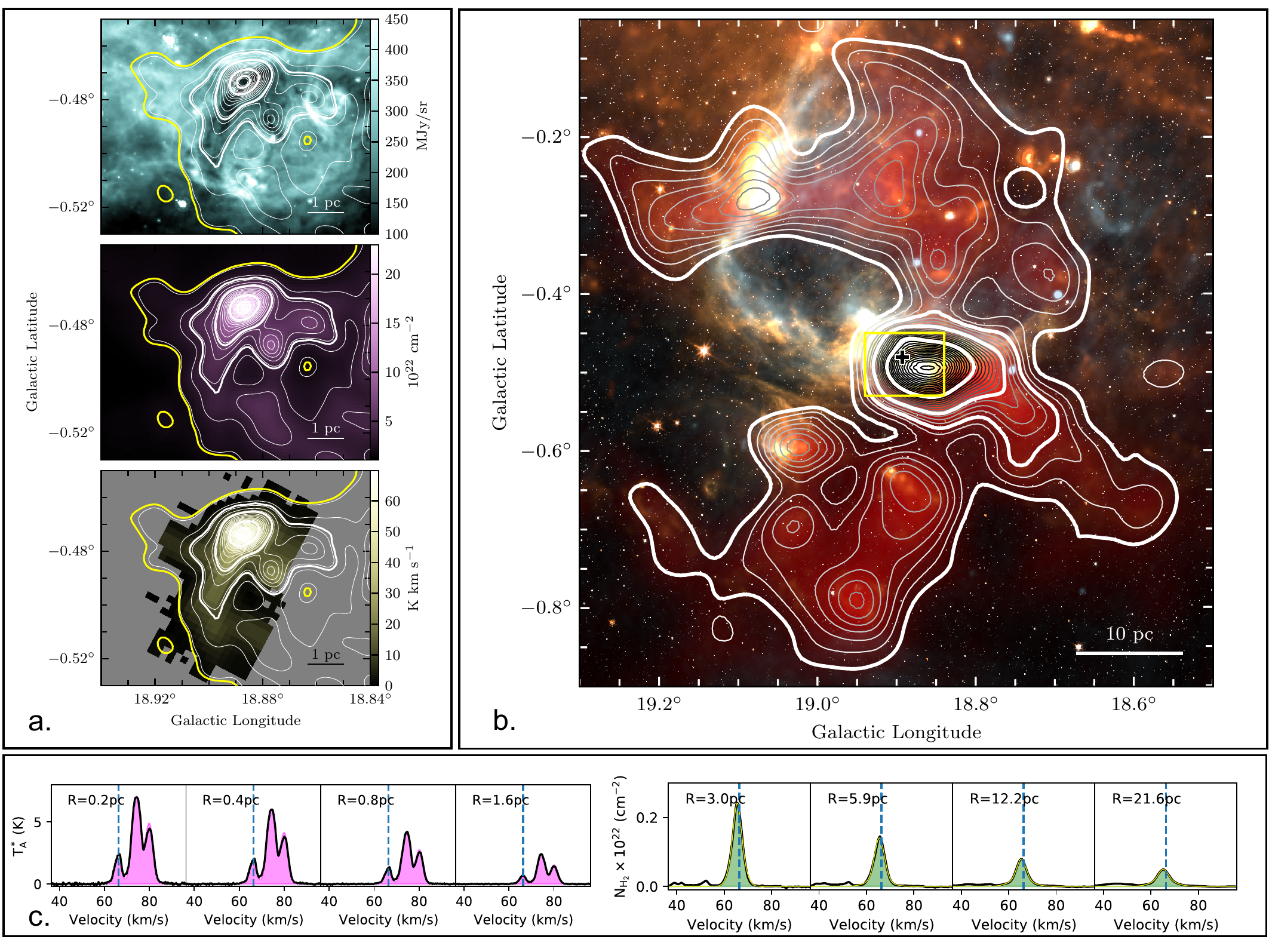}
      \caption{Images of SDC18.888-0.476. (a): top - {\it Spitzer} 8\micron; middle- H$_2$ column density from {\it Herschel} observations; bottom - N$_2$H$^+$(1-0) integrated emission. The contours are identical in all panels, and are those of the H$_2$ column density image. The yellow contour corresponds to $N_{\rm{N_2H^+}}^{\rm{edge}}$. The four thicker white contours are those used to compute the average N$_2$H$^+$(1-0) spectra displayed in magenta in panel, the first one of which corresponds to $N_{\rm{N_2H^+}}^{\rm{start}}$. (c). (b): Multi-colour image of the molecular cloud hosting the SDC18.888-0.476 infrared dark clump (white: 3.6\,\micron, orange: 8\,\micron, yellow: 70\,\micron, orange: 350\,\micron, blue: 1.42\,GHz, red: H$_2$ column density). The contours show the H$_2$ column density obtained from the Galactic Ring Survey $^{13}$CO(1-0) data. The thicker white contours are those used to compute the $^{13}$CO(1-0)-based spectra shown in green in panel (c). The plus symbol shows the central position of the IRDC, and the yellow rectangle shows the coverage of the images displayed in (a). (c): Spectra averaged within the highlighted H$_2$ column density contours in panels (a) and (b). The radius of the region within which the spectra have been averaged are indicated in each panel. The vertical blue dashed lines show the systematic clump velocity as measured from \n2hp . The compilation of the data presented in this figure summarises all the information used for each cloud in the study presented here. A similar figure for each remaining IRDC can be found in Appendix A. }
         \label{example}
   \end{figure*}

A large number of studies have looked at the gravitational binding of molecular clouds and their sub-structures within, most often via the calculation of their virial parameters \citep[e.g.][]{larson1981, solomon1987, heyer2009, roman-duval2010, kauffmann2013, schuller2017, miville-deschenes2017, rigby2019,duarte-cabral2021}. Depending on the cloud sample that is being studied, the methods that are being used, and the interpretation  of the data that is being made, the conclusions range from molecular clouds are: in hydro-static equilibrium, collapsing, or unbound. As a result, a consensus as yet to be found.

 A possibly more insightful analysis of molecular clouds is the study of their internal virial ratio profiles. Indeed, if there is a scale/density threshold at which the gravitational binding of clouds change from unbound to bound as a result of, for instance, stellar feedback, then the virial ratio profiles of individual clouds should exhibit some breaks at that particular scale. While  several studies have  investigated the shape of the mass profiles of cores, clumps and clouds \citep[e.g.][]{motte2001,kauffmann2010b,palau2014,barnes2021}, studies that have looked into their virial ratio profiles are a lot more rare. This is the consequence of the much larger range of spatial scales probed by dust observations, most often used for structure mass estimates, compared to spectral line ones, needed to derive velocity dispersions of those structures. The few observational studies that have looked into the question of clouds' virial ratio radial profiles have done so either on single clouds \citep[e.g.][]{rosolowsky2008,goodman2009}, using single tracers \citep[e.g.][]{heyer2009, li2015, wong2019}, or using only two radial points \citep[e.g.][]{heyer2009,traficante2018,traficante2020}. As a result, the virial ratio radial profiles of molecular clouds have not been fully characterised yet. 
  In this paper,  we use a multi-scale and multi-tracer approach that allows us to construct, in a uniform way, the virial ratio profiles of a sample of molecular clouds from scales of tenths of a parsec up to scales of tens of parsecs. In Sec. 2 we present the source selection and observations. Section 3 explains how the profiles of individual cloud are built. Section 4 presents the models we use to determine the origin of the observed profile features. In Sec. 5  we discuss our results while conclusions are laid out in Sec.6. \\

\section{Source selection and observations}

\subsection{Sample}

We selected a sample of 27 infrared dark clouds (IRDCs) from the Spitzer Dark Cloud catalogue of \citet{peretto2009}. Compared to other cloud samples, IRDCs have the advantage that their heliocentric distances are better constrained, with a large majority of IRDCs lying at the near kinematic distance solution provided by Galactic rotation models \citep{ellsworth-bowers2013}. In this paper, the adopted distances for all IRDCs are the near kinematic distance solutions from the \citet{reid2009} model.  The selection criteria for these IRDCs are: (a) The kinematic distance as estimated from $^{13}\rm{CO}(1-0)$ GRS data \citep{roman-duval2010} should be $d=4(\pm1)$~kpc; (b) Selected IRDCs should exhibit a range of aspect ratios, i.e. from circular to filamentary, as measured from {\it Herschel} column density images \citep{peretto2016}; (c) Selected IRDCs should exhibit a range of mass and size as estimated from Herschel column density images; (d) all IRDCs have to lie beyond $l=15\degr$ in order to be easily observed from the IRAM 30m telescope. Global properties of the 27 selected clouds can be found in Table 1. Note that kinematic distances have been recalculated using the dense gas data presented in this paper, leading in a few cases to a departure from condition (a). Figure~\ref{example}a shows one of the selected IRDCs, images of the remaining 26 can be seen in Appendix A  which is supplied as online supplementary material.

\begin{center}
\begin{table*}
\caption{Infrared dark cloud sample}             
\label{table:1}      
\centering                          
\begin{tabular}{c c c c c}        
\hline\hline                 
Cloud ID & Name & Coordinates & Systemic velocity& Distance  \\    
& & (J2000) & (km/s) & (kpc)  \\
\hline                        
1 & SDC18.624-0.070 & 18:25:10.0 -12:43:45 & +45.6				& 3.50 \\
2 & SDC18.787-0.286 & 18:26:19.0 -12:41:16 & +65.4  				& 4.36  \\
3 & SDC18.888-0.476 & 18:27:09.7 -12:41:32 & +66.3   				& 4.38 \\
4 & SDC21.321-0.139 & 18:30:32.1 -10:22:50 & +66.5   				&4.24  \\
5 & SDC22.373+0.446 & 18:30:24.5 -09:10:34 & +53.0  				& 3.61 \\
6 & SDC22.724-0.269 & 18:33:38.3 -09:11:55 & +73.3 (+105.0) 		& 4.44 \\
7 & SDC23.066+0.049 & 18:33:08.2 -08:44:53 & +91.8 				& 5.11 \\
8 & SDC23.367-0.288 & 18:34:53.8 -08:38:00 & +78.3 (+103.0; +58) 	& 4.60  \\
9 & SDC24.118-0.175 & 18:35:52.6 -07:55:06 & +80.9 				&4.68  \\
10 & SDC24.433-0.231 & 18:36:41.0 -07:39:20 & +58.4 				& 3.75 \\
11 & SDC24.489-0.689 & 18:38:25.7 -07:49:36 & +48.1 				&3.28   \\
12 & SDC24.618-0.323 & 18:37:22.4 -07:32:18 & +43.4 				&3.04  \\
13 & SDC24.630+0.151 & 18:35:38.2 -07:18:35 & +53.2 (+115.0)		&3.51 \\
14 & SDC25.166-0.306 & 18:38:13.0 -07:03:00 & +63.6 				&3.95 \\
15 & SDC25.243-0.447 & 18:38:57.1 -07:02:20 & +59.1 				&3.75 \\
16 & SDC26.507+0.716 & 18:37:07.9 -05:23:58 & +48.3 				& 3.21  \\
17 & SDC28.275-0.163 & 18:43:30.3 -04:12:45  & +80.3				&4.60  \\
18 & SDC28.333+0.063 & 18:42:54.1 -04:02:30 & +79.3 				&4.56  \\
19 & SDC31.039+0.241 & 18:47:03.3 -01:33:50 & +78.2 (+98; +110) 	& 4.54\\
20 & SDC34.370+0.203 & 18:53:18.9 +01:24:54 & +57.9	 			& 3.59\\
21 & SDC35.429+0.138 & 18:55:30.4 +02:17:10 & +77.0 				&4.67  \\
22 & SDC35.527-0.269 & 18:57:08.6 +02:09:08 & + 45.4  			&  2.95  \\
23 & SDC35.745+0.147 & 18:56:02.6 +02:34:44 & +83.4 				&5.11   \\
24 & SDC38.850-0.427 & 19:03:46.8 +05:04:03 & +42.2  				&2.81  \\
25 & SDC40.283-0.216 & 19:05:41.2 +06:26:09 &  +72.7 				& 4.89 \\
26 & SDC47.061+0.257 & 19:16:41.8 +12:39:39 & +57.0  			& 4.64  \\
27 & SDC52.723+0.045 & 19:28:34.4 +17:34:17 & +44.1  			&4.49   \\
\hline                                   
\end{tabular}
\begin{tablenotes}
      \small
      \item Col.~1: IRDC identification number; Col.~2: IRDC name from \cite{peretto2009}; Col.~3: central IRDC coordinates; Col.~4: Systemic LSR velocity of the clump as estimated from N$_2$H$^+$(1-0), the velocities in between brackets correspond to the additional components identified in the spectra; Col.~5: Near kinematic distance as estimated from the \cite{reid2009} model, uncertainties on those are typically of the order of 10\% to 20\%. 
    \end{tablenotes}
\end{table*}
 \end{center}

\subsection{Observations}
In this study, we exploit four different datasets, each of which is tracing a specific density regime of molecular clouds and/or giving us access to different sets of information (mass versus kinematics). In the following, we describe each of these datasets.

\subsubsection{ N$_2$H$^+$(1-0) data}
We observed the 27 infrared dark clouds at the IRAM 30m between the 18th and 24th of June 2013, reaching a total of 42h of telescope time. The weather conditions were stable with an average sky opacity at 230~GHz of 0.2. We mapped each region using the 90~Hz EMIR receiver  in conjunction with the FTS spectrometer at 50~kHz spectral resolution, providing a velocity resolution of 0.16~km/s.  Primary pointing and focus were performed on Saturn. The pointing accuracy was $<5\arcsec$. In this study we focus on the N$_2$H$^+$(1-0)  line, with  an angular resolution of 28\arcsec\ . All data have been reduced using the CLASS package, and gridded into 9\arcsec\ pixel-size cubes. The final noise range from 0.09~K to 0.2~K per velocity channel and pixel.

\subsubsection{{\it Herschel} data}

We used the PACS \citep{poglitsch2010} and SPIRE \citep{griffin2010}  {\it Herschel} \citep{pilbratt2010} data from the Hi-GAL survey \citep{molinari2010}. 
The Hi-GAL data were reduced, as described in \cite{traficante2011}, using HIPE \citep{ott2010} for calibration and deglitching (SPIRE only), routines especially developed for Hi-GAL data reduction (drift removal, deglitching), and the ROMAGAL map-making algorithm. Post-processing on the maps was applied to help with image artefact removal \citep{piazzo2015}. In this paper, we make use of  the PACS 160~\micron\ and SPIRE 250/450/500 \micron\ data with a nominal angular resolution of 12\arcsec, 18\arcsec, 25\arcsec, and 36\arcsec, respectively.
In addition, zero-flux levels for every Hi-GAL field have been recovered by correlating Herschel data with Planck and IRAS data \citep{bernard2010}.

\subsubsection{$^{13}$CO(1-0) and $^{12}$CO(1-0) data}

We used the FCRAO $^{13}$CO(1-0) data from the Galactic Ring Survey \citep[GRS -][]{jackson2006} along with the FCRAO UMSB $^{12}$CO(1-0) data \citep{sanders1986, clemens1986}. The GRS data has an angular resolution of 44\arcsec, a velocity resolution of 0.21 km/s and a one $\sigma$ noise of 0.13K (in $T_A^*$ scale). The main beam efficiency of the FCRAO telescope at the $^{13}$CO(1-0) frequency is 0.48. All clouds from our sample of 27 IRDCs are covered by the GRS.  

The UMSB $^{12}$CO(1-0) data has a nominal angular resolution of 44\arcsec. However, the data has been sampled on a  3\arcmin-grid, which effectively decreases the resolution.  The velocity resolution is 1~km/s, and the one $\sigma$ noise is 0.4K (in $T_R^*$) scale. In order to be able to convert that into a main beam temperature one needs first to multiply  by $n_{ffs}=0.7$  which converts the unit back to $T_A^*$  \citep{kutner1981, sanders1986} and then divide by the main beam efficiency 0.48, so effectively multiplying the UMSB dataset by a (0.7/0.48) factor.

   \begin{figure}
   \centering
   \includegraphics[width=8.cm]{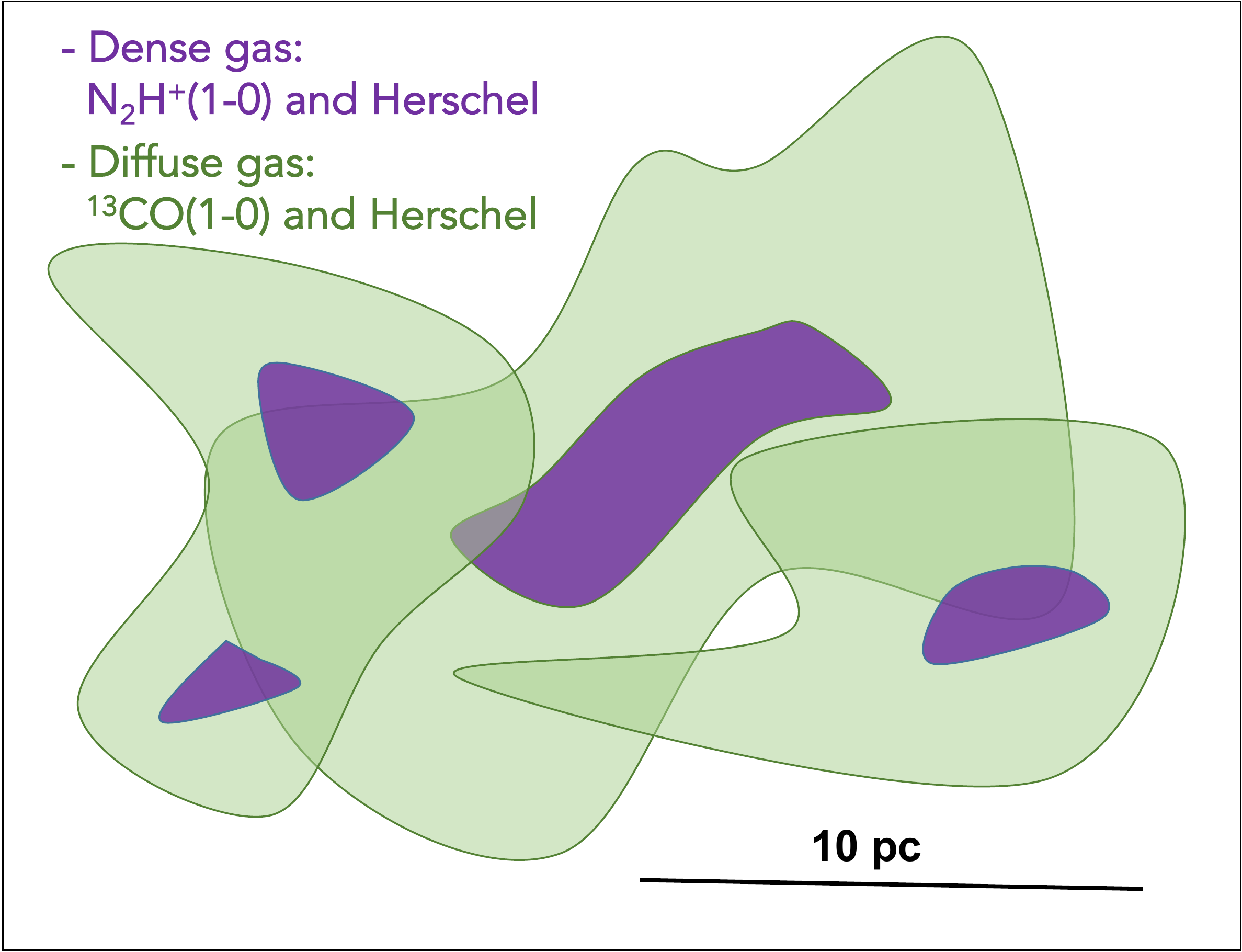}
      \caption{Sketch of molecular cloud configuration and relevant tracers. Diffuse gas (represented in green) is traced by  $^{13}$CO(1-0) and dust continuum. However, only the former is able to disentangle the emission of multiple clouds along the line of sight by segmenting them in velocity space. In this paper, we will use both tracers to constrain the mass and morphology of the clouds on the largest scales. Dense gas (represented in purple) is well probed by both dust continuum and molecular line tracers such as N$_2$H$^+$(1-0). It is very rare that two N$_2$H$^+$(1-0) cloud overlap (as the low frequency of multiple N$_2$H$^+$(1-0) velocity components is showing). Dust continuum can therefore also be used once a background contamination (from the diffuse gas) has been removed.} 
         \label{sketch}
   \end{figure}

\section{Mass and velocity dispersion profiles}

The goal of this paper is to determine how the ratio of kinetic to gravitational energy of clouds changes as a function of spatial scale. In order to observationally measure such ratio, one needs to determine three quantities: radius, mass, and velocity dispersion. While the cloud mass can reliably be determined via dust emission observations, no single molecular line can trace molecular gas velocity dispersion on all scales, either because of high optical depth or low abundance. We therefore need a combination of tracers to trace different parts of the cloud. Here, we use $^{13}$CO(1-0) to trace the large scales, more diffuse parts of the clouds, and N$_2$H$^+$(1-0)  to trace their densest parts. Figure~\ref{sketch} shows a simple sketch that illustrates what tracer we use for what purpose. In the following subsections we describe how we computed the three required quantities for both the dense and diffuse regions of the clouds.

\subsection{Dense gas}

\subsubsection{{\it Herschel} column density maps of IRDCs}

For the purpose of this study, we  computed H$_2$ column density maps using the method presented in \citet[referred to as P16 hereafter]{peretto2016}. That method consists in using the ratio of the {\it Herschel} 160~\micron\ over 250~\micron\ dust emission to measure the temperature of the dust, and then use it, in combination with the 250\micron\ image to derive the column density of gas (assuming a dust to gas mass ratio of 1\%) at an angular resolution of 18". For the purpose of the study presented here, we  convolved the column density image to the same angular resolution as the N$_2$H$^+$(1-0) data, i.e. 28".  The assumed specific dust opacity is $\kappa_{{\lambda}}=0.1\left(\frac{\lambda}{300\micron}\right)^{\beta}$~cm$^2$\,g$^{-1}$ \citep{hildebrand1983}, with $\beta=1.8$  \citep[e.g.][]{planck2011, sadavoy2016,rigby2018}.

When computing these maps, we make the assumption of a uniform temperature along the line of sight. This is of course incorrect but it is not completely clear though how wrong this assumption is for the structures we are studying. Since we might expect this assumption of a single temperature to be the most inaccurate towards the centre of each clump, we decided to compare the mass profiles of each clumps obtained with P16's method with that of PPMAP \citep{marsh2015}, a bayesian code that derive, from {\it Herschel} observations, the distribution of dust temperatures along the line of sight. Note that we do not use PPMAP in this paper as it can generate a number of artefacts around bright protostellar sources, it is computationally expensive, and  arising issues are a lot less straightforward to identify than when using the P16's method.

\begin{figure}
   \centering
   \includegraphics[width=9.cm]{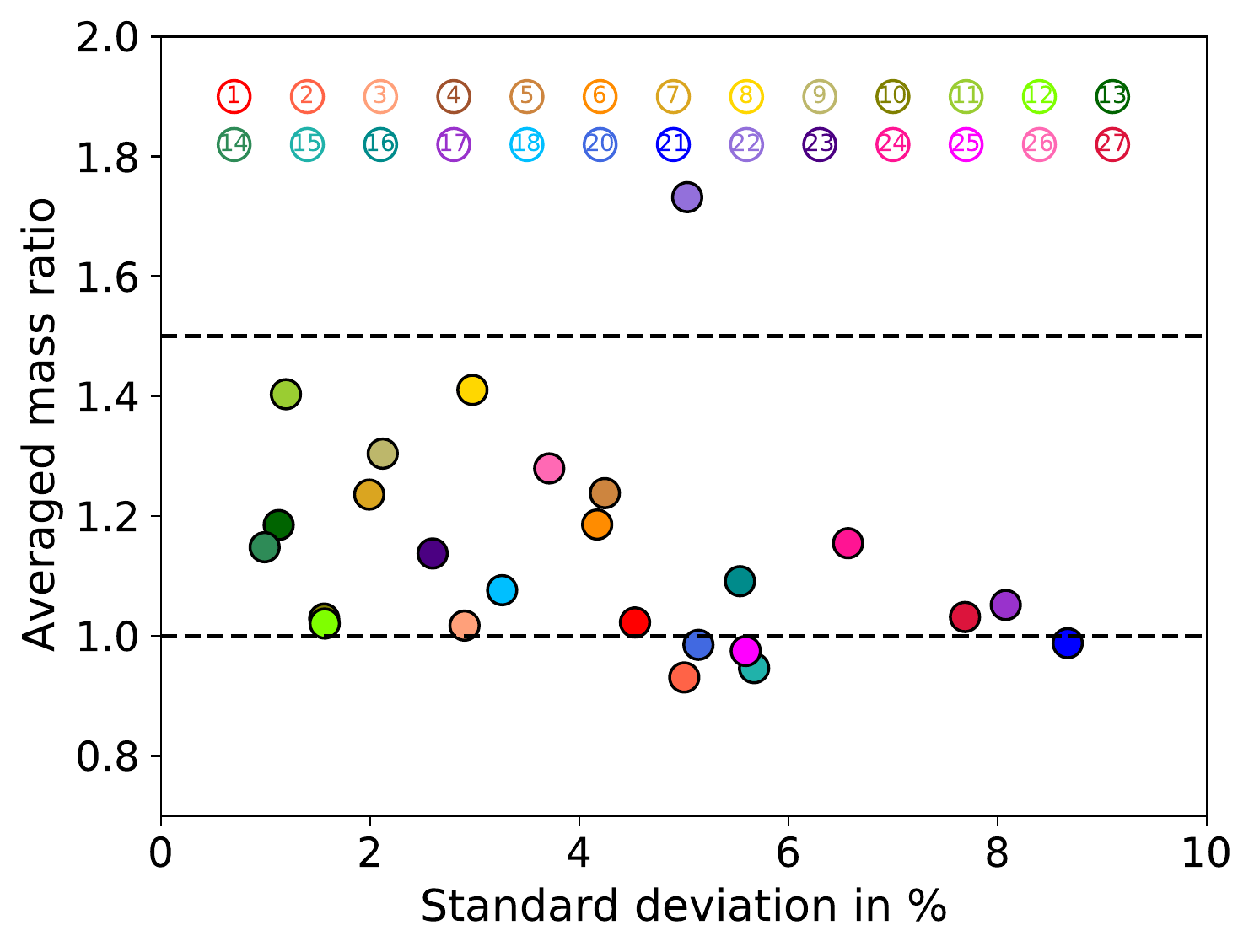} 
      \vspace{-0.5cm}
      \caption{Ratio of the PPMAP masses over the P16 masses averaged over their radial profiles as a function of their mass ratio standard deviation. The black dashed lines  show  mass ratios of 1 and 1.5. Each colour corresponds to a single IRDC whose ID number can be found at the top of the figure (see Table 1 for the corresponding IRDC name). Note that IRDC SDC31.039+0.241 (ID number 19) has been left out as a result of the presence of multiple dense clumps present along the line-of-sight (see Sec. 3.1.2).}
         \label{massratio}
   \end{figure}

On the y-axis of Fig.~\ref{massratio} we show the ratio of the PPMAP over the P16 masses, radially averaged. On the x-axis of the same figure we show the radial dispersion of that same ratio, i.e. how much it varies about the average value as a function of radius (i.e. 0\% means that the ratio is radially uniform). One can see that while, on average, the PPMAP masses are about 20\% larger than the P16 masses, the variations of the mass ratio as a function of radius are small, and remain below 5\% for most clouds, with a maximum standard deviation of less than 8\%. This shows that, while there might be a systematic uncertainty on the mass of 20\%, the shape of the mass profiles derived from both methods are very much consistent with each other.

 \subsubsection{N$_2$H$^+$(1-0) as a tracer of {\it Herschel} clumps}

All 27 IRDCs are detected in \n2hp.  For 4 of them ($\sim 15\%$), multiple clouds with velocities differing by more than 20~km/s have been identified within the observed field of views. For one of this IRDC (SDC31.039+0.241), the \n2hp\ emission of the different clouds spatially overlap. This cloud is therefore excluded from the rest of the analysis as the origin of the corresponding dust continuum emission becomes very uncertain. Regarding the remaining three clouds (SDC22.724-0.269, SDC23.367-0.288, SDC24.630+0.151), we only consider  the cloud for which the \n2hp\ integrated emission best matches the extinction feature seen in the mid-infrared. The corresponding velocities are provided in Table 1. 

 Another 4 IRDCs (SDC24.433-0.231, SDC24.630+0.151, SDC26.507+0.716, SDC35.527-0.269) show multiple velocity components with velocity differences lower than 3 km/s, only one of these also exhibits multiple clouds along the line of sight (SDC24.630+0.151). However, once averaged within column density contours (see Appendix A), the multiple velocity components are mostly washed out, and are therefore not a concern in the context of this study. Note that one of the multiple velocity component cloud, i.e. SDC35.527-0.269,  has been extensively studied in the past at high angular resolution clearly revealing multiple velocity component structures \citep[e.g.][]{henshaw2014}.
  
\begin{figure}
   \centering
   \includegraphics[width=9.cm]{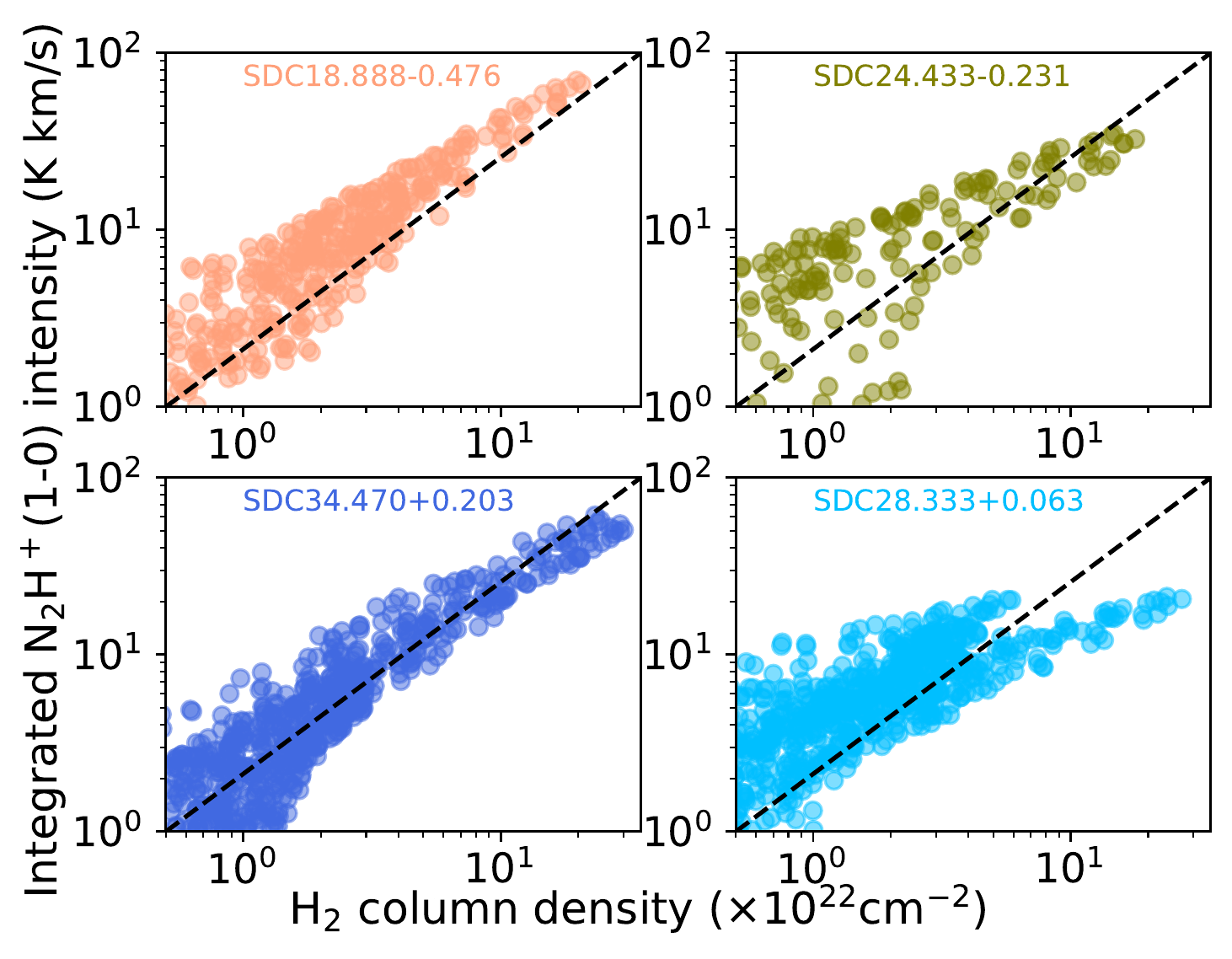}
   \vspace{-0.5cm}
      \caption{Scatter plots of the background subtracted  H$_2$ column density versus the \n2hp\ integrated intensity (in T$_a^*$ scale) for four clumps. In each panel the same linear relation is displayed as a dashed black line. }
         \label{scatter}
   \end{figure}

The  morphology of the \n2hp\ integrated intensity images are very similar to that of the H$_2$ {\it Herschel} column density maps (see Fig.~\ref{example}), qualitatively showing that  \n2hp\ is a good tracer of the column density structure of star-forming clouds. In order to quantify the correlation between dust column density and \n2hp\  line emission we produced scatter plots for each cloud of the H$_2$ column density derived from {\it Herschel}, for which the background as defined by N$_{\rm{N_2H^+}}^{\rm{edge}}$ (see next section) has been subtracted,  versus the integrated intensity of \n2hp (see Fig,~\ref{scatter} for four representative examples). One can see there is, indeed, a strong linear correlation between the two quantities, with only small departures from it for some clouds exhibiting a large range of different physical conditions  (see the case of SDC28.333+0.063 in Fig.~\ref{scatter}). We observe similar correlations for all clouds for which there is enough dynamic range and independent points (i.e. 21/27 clouds). We have also checked whether the relation provided by \citet{hacar2018} between H$_2$ column density, \n2hp\ integrated intensities, and temperature hold for our cloud sample. We can confirm that it does for most of the clumps, but some significant departures are observed, which can be explained by a variation of the N$_2$H$^+$ abundance by a factor of 2 or so.  Nevertheless, from this comparison we can conclude that N$_2$H$^+$ is a good tracer of the dense gas as traced with {\it Herschel}, and therefore that we can reasonably use it to trace the  kinematics of {\it Herschel} clumps for the (column) density range we are probing (i.e. N$_{\rm{H_2}}\ge 10^{22}$~cm$^{-2}$). As such we do not expect the effect of using different tracers for mass and kinematics to be a significant issue in our study \citep[see][]{traficante2018,yuan2020}.

 \subsubsection{Mass and velocity dispersion estimates}
 The resulting H$_2$ column density maps (see Fig.~\ref{example}) are contaminated by foreground and background interstellar structures which are not physically associated to the cloud. Removing such contributions is not an easy task \citep{peretto2010b,battersby2011}. In the context of this study, we are mostly interested in the part of the cloud which is seen in \n2hp\ in the IRAM 30m data.  Therefore we define the "edge" of the dense part of the clouds as being the column density contour, $N_{\rm{N_2H^+}}^{\rm{edge}}$,  that best matches the extent of the  \n2hp\  integrated intensity map. This is done by computing the median (along with the 16$^{\rm{th}}$ and 84$^{\rm{th}}$ percentiles) column density value within a ring just outside the N$_2$H$^+$(1-0) integrated intensity contour of 0.5~K\,km/s, i.e. our detection limit. The value of $N_{\rm{N_2H^+}}^{\rm{edge}}$ will then serve as the background column density of the clump that we will remove from any clump scale mass measurements (see Table~\ref{table:2} for the individual values of $N_{\rm{N_2H^+}}^{\rm{edge}}$ and corresponding 16$^{\rm{th}}$ and 84$^{\rm{th}}$ percentiles).

We used the contour-based dendrogram tool from  \citet{peretto2009}  on the {\it Herschel} column density maps to estimate sizes and masses of connected groups of pixels lying above a certain column density. In order to be considered for the analysis those groups of pixels need to be larger than the number of pixels within an angular resolution element, and need to be part of a structure whose column density amplitude from local maximum to local minimum is larger than a predefined threshold, $N_{\rm{H_2}}^{\rm{th}}$. The column density increment we used in our dendrogram analysis is  $\sigma_{N_{\rm{H_2}}}=2\times10^{21}$~cm$^{-2}$ for all clouds, with  $N_{\rm{H_2}}^{\rm{th}}= 5\sigma_{N_{\rm{H_2}}}$. The starting column density contour, $N_{\rm{N_2H^+}}^{\rm{start}}$ (see Table~\ref{table:2}) is set to be larger or equal to $N_{\rm{N_2H^+}}^{\rm{edge}}$, and is determined by eye. The reason for not systematically having $N_{\rm{N_2H^+}}^{\rm{start}}=N_{\rm{N_2H^+}}^{\rm{edge}}$ is that the $N_{\rm{N_2H^+}}^{\rm{edge}}$ contour can be more extended than the coverage of our \n2hp\ maps, and therefore, in such cases, the computed masses would be overestimated. The mass of any identified group of pixels is then given by:

\begin{equation}
M_{\rm{N_2H^+}}=\Omega_{\rm{pix}}^{\rm{N_2H^+}}d^2\mu_{\rm{mol}}m_{\rm{H}}\sum_{i=1}^{n_{\rm{pix}}^{\rm{N_2H^+}}}\left(N_{\rm{H_2},i}-N_{\rm{N_2H^+}}^{\rm{edge}}\right)
\label{clumpmass}
\end{equation}

\noindent where the sum is on all the $n_{\rm{pix}}^{\rm{N_2H^+}}$ pixels belonging to the group of interest, $\Omega_{\rm{pix}}^{\rm{N_2H^+}}$ is the solid angle subtended by a pixel, $d$ is the distance to the IRDC, $\mu_{\rm{mol}}$ is the mean molecular weight and is set to 2.33,  and $m_{\rm{H}}$ is the mass of the hydrogen atom. The estimated radius associated to that group is calculated via the following relation:

\begin{equation}
R_{\rm{N_2H^+}}=\sqrt{\frac{n_{\rm{pix}}^{\rm{N_2H^+}}\Omega_{\rm{pix}}^{\rm{N_2H^+}}d^2}{\pi}}
\end{equation}

\noindent  Projected masses estimated this way will always overestimate the mass enclosed within the volume of  radius $r$ as lower-density material along the line-of-sight is wrongly associated to that volume (see Sec. 4). In Table~\ref{table:2} we provide the radius, mass, and aspect ratio of each IRDC at the starting column density contour $N_{\rm{N_2H^+}}^{\rm{start}}$. Note that we also give the mass uncertainties related to the $16^{\rm{th}}-84^{\rm{th}}$ percentiles range of $N_{\rm{N_2H^+}}^{\rm{edge}}$ values.

In order to estimate the dense gas velocity dispersion of the structures identified in the dendrogram of the {\it Herschel} column density images, we computed their corresponding \n2hp\ spectra, averaged over all $n_{\rm{pix}}^{\rm{N_2H^+}}$ pixels that belong to the relevant group. We then used a python routine, inspired from that of GILDAS/CLASS, that uses the {\it curvefit} minimisation routine in order to fit the 7 hyperfine components of the \n2hp\ transition. The parameters of the fit are: the central velocity $\varv_{\rm{N_2H^+}}$, the velocity dispersion of the gas $\sigma_{\rm{N_2H^+}}$, the sum of the central opacities of the 7 hyperfine components $\tau_{\rm{N_2H^+}}$, and the sum of the antenna temperature peaks of the 7 hyperfine components $T_{\rm{ant}}^{\rm{tot}}$. The velocity dispersion $\sigma_{\rm{N_2H^+}}$ is obtained from the best fitting model. Examples of average  \n2hp\  and their fit are displayed in Fig.~\ref{example} for SDC18.888 (and in Appendix A for the other clouds).  We fitted only one velocity component to all spectra, even for those showing potential multiple velocity components as those are almost systematically blended once the emission is averaged within each contour. The velocity dispersions $\sigma_{\rm{N_2H^+}}^{\rm{start}}$ estimated within the column density contour $N_{\rm{N_2H^+}}^{\rm{start}}$ are given in Table~\ref{table:2}.\\

\begin{table*}
\caption{IRDCs and parent cloud properties}             
\label{table:2}      
\centering                          
\begin{tabular}{c c c c r c c r r c c}        
\hline\hline                 
ID & $N_{\rm{N_2H^+}}^{\rm{edge}}$ &  $N_{\rm{N_2H^+}}^{\rm{start}}$ &  $R_{\rm{N_2H^+}}^{\rm{start}}$ & $M_{\rm{N_2H^+}}^{\rm{start}}$ &  $\sigma_{\rm{N_2H^+}}^{\rm{start}}$ &  $\alpha_{\rm{vir}}^{\rm{N_2H^+}}$ &$R_{\rm{^{13}CO}}^{\rm{start}}$ & $M_{\rm{^{13}CO}}^{\rm{start}}$ &  $\sigma_{\rm{^{13}CO}}^{\rm{start}}$ & $\alpha_{\rm{vir}}^{\rm{^{13}CO}}$\\    
&($\times 10^{22}$~cm$^{-2}$) &($\times 10^{22}$~cm$^{-2}$)& (pc) & (M$_{\odot}$) & 	(km/s)  & & (pc) & ($\times10^4$~M$_{\odot}$) & (km/s) &\\
\hline                        
1  & 		$2.8^{+0.3}_{-0.2}$		& 3.3 		& 1.66				& $3239^{+365}_{-546}$  	&	0.99	 &	0.62		& 12.62				& 14.52  	&	5.74	&	3.35\\
2  & 		$2.4^{+0.4}_{-0.4}$ 		& 3.3 		& 1.18  				& $1807^{+365}_{-365}$  	&	1.06	&	0.89		& 21.54				& 48.36  	&	4.19	&	0.91\\
3  & 		$1.9^{+0.3}_{-0.3}$ 		& 3.7 		& 1.61   				& $7882^{+513}_{-513}$	 &	1.55	&	0.58		& 21.63				& 47.30  	&	3.72	&	0.74\\
4  & 		$3.2^{+0.3}_{-0.5}$ 		& 3.3 		& 1.09   				& $562^{+395}_{-237}$	 &	0.69	&	1.20		& 9.02				& 3.71  	&	2.81	&	2.25\\
5  & 		$2.3^{+0.4}_{-0.5}$ 		& 2.9			& 0.63  				& $437^{+101}_{-81}$	& 	1.13	&	2.03		& 3.46				& 0.31  	&	1.39	&	2.61\\
6  & 		$3.3^{+0.7}_{-0.5}$ 		& 4.6 		& 0.60				& $533^{+118}_{-164}$ 	&	0.65	&	0.62		& 11.58				& 11.85  	&	4.50	&	2.32\\
7  & 		$3.2^{+0.4}_{-0.3}$		& 3.6 		& 1.48 				& $1536^{+434}_{-579}$  	&	0.98	&	1.14		& 16.27				& 30.40  	&	5.90	&	2.18\\
8  & 		$4.2^{+0.5}_{-0.5}$ 		& 6.1 		& 0.72				& $1575^{+172}_{-172}$	&	1.17	&	0.76		& 17.76				& 27.61  	&	3.67	&	1.02\\
9  &		$2.6^{+0.3}_{-0.4}$ 		& 2.9 		& 1.32 				& $1318^{+462}_{-346}$	 &	1.03	&	1.31		& 6.46				& 2.02  	&	1.63	&	1.00\\
10 & 		$2.9^{+0.5}_{-0.5}$ 		& 3.1 		& 1.39 				& $3282^{+642}_{-642}$	 &	1.50	&	1.14		& 5.23				& 2.31  	&	2.88	&	2.21\\
11 &		$1.2^{+0.2}_{-0.2}$ 		& 1.4 		& 1.12				& $831^{+167}_{-167}$   	&	1.13	&	2.08		& 9.19				& 2.18  	&	1.67	&	1.39\\
12 & 		$2.0^{+0.3}_{-0.1}$ 		& 2.5 		& 0.82 				& $594^{+45}_{-134}$   	&	0.85	&	1.26		& 12.67				& 5.85  	&	1.63	&	0.68\\
13 & 		$2.7^{+0.5}_{-0.3}$ 		& 4.4 		& 0.48				& $472^{+45}_{-75}$ 	&	1.08	&	1.42		& 6.80				& 3.34  	&	3.58	&	3.06\\
14 & 		$2.2^{+0.2}_{-0.2}$ 		& 2.5			& 1.68 				&$3622^{+374}_{-374}$ 	 &	0.96	&	0.53		& 8.65				& 3.23  	&	1.78	&	1.00\\
15 &		$1.8^{+0.2}_{-0.2}$ 		& 2.0			& 1.33 				&$1233^{+235}_{-235}$ 	 &	0.84	&	0.95		& 13.11				& 13.35  	&	4.23	&	2.05\\
16 &		$1.7^{+0.3}_{-0.3}$ 		& 1.8 		& 1.62 				& $1474^{+518}_{-518}$	  &	1.02	&	1.40		& 7.28				& 3.20  	&	1.44	&	0.56\\
17 &		$3.0^{+0.5}_{-0.3}$		 & 3.7 		& 1.53				&$2687^{+465}_{-775}$	  &	1.27	&	1.10		& 31.13				& 98.02  	&	4.31	&	0.69\\
18 & 		$4.4^{+0.4}_{-0.4}$ 		& 5.1 		& 2.65 				&$13954^{+1863}_{-1863}$ &	1.42	&	0.46		& 30.85				&103.70  	&	5.27	&	0.97\\
20 & 		$2.7^{+0.5}_{-0.4}$ 		& 3.5 		& 2.15	 			& $12048^{+1223}_{-1528}$&	1.34	&	0.39		& 11.33				& 18.13  	&	3.79	&	1.05\\
21 & 		$2.7^{+0.4}_{-0.3}$ 		& 3.0 		& 2.62 				&$7761^{+1366}_{-2276}$  &	1.33	&	0.72		& 11.24				& 4.01  	&	3.20	&	3.37\\
22 & 		$2.2^{+0.3}_{-0.2}$ 		& 2.7 		& 1.36  				&  $1499^{+245}_{-367}$	  &	0.77	&	0.68		& 8.22				& 3.49  	&	1.89	&	0.99\\
23 & 		$1.9^{+0.1}_{-0.2}$ 		& 2.0 		& 2.01  				&  $3490^{+536}_{-268}$	  &	1.07	&	0.80		& 11.58				& 7.66  	&	3.74	&	2.47\\
24 &		$2.7^{+0.2}_{-0.2}$ 		& 2.9 		&  0.77				& $276^{+79}_{-79}$   	&	0.62	&	1.42		& 8.46				& 4.29  	&	1.52	&	0.55\\
25 & 		$2.1^{+0.2}_{-0.2}$ 		& 2.2			&  1.81 				& $4044^{+432}_{-432}$ 	 &	1.47	&	1.15		& 11.06				& 5.00  	&	1.32	&	0.46\\
26 & 		$1.7^{+0.5}_{-0.5}$ 		& 2.2 		& 1.35  				& $1982^{+601}_{-601}$   	&	1.29	&	1.36		& 13.34			& 12.67  	&	2.48	&	0.76\\
27 & 		$1.7^{+0.2}_{-0.2}$ 		& 1.7 		& 1.37  				& $358^{+247}_{-247}$   	 &	0.67	&	2.20		& 17.31				& 16.24  	&	2.37	&	0.71\\
\hline                                   
\end{tabular}
\begin{tablenotes}
      \small
      \item Col.~1: IRDC identification number; Col.~2: H$_2$ column density matching the edge of the $N_2$H$^+$(1-0) emission (median, 16$^{\rm{th}}$ and 84$^{\rm{th}}$ percentiles); Col.~3: H$_2$ column density from which the dendrogram tree starts;  Col.~4: IRDC radius corresponding to  $N_{\rm{N_2H^+}}^{\rm{start}}$; Col.~5: Gas mass within $R_{\rm{N_2H^+}}^{\rm{start}}$, uncertainties reflect the mass changes when considering the 16$^{\rm{th}}$ and 84$^{\rm{th}}$ percentiles of $N_{\rm{N_2H^+}}^{\rm{edge}}$; Col.~6: N$_2$H$^+$(1-0) velocity dispersion estimated within $R_{\rm{N_2H^+}}^{\rm{start}}$; Col.~7: virial ratio estimated within  $R_{\rm{N_2H^+}}^{\rm{start}}$;  Col.~8: Parent cloud radius; Col.~9: Gas mass within $R_{\rm{^{13}CO}}^{\rm{start}}$; Col.~10: $^{13}$CO(1-0) velocity dispersion estimated within  $R_{\rm{^{13}CO}}^{\rm{start}}$; Col.~11: virial ratio estimated within $R_{\rm{^{13}CO}}^{\rm{start}}$
    \end{tablenotes}
\end{table*}

\subsection{Diffuse gas}

One key aspect of this study is to estimate cloud properties on a large range of scales. While we presented in the previous section how we estimate the properties of the clumps, we want now to connect these to their more diffuse envelopes. To achieve this we use a combination of the {\it Herschel} dust continuum data and the $^{13}$CO(1-0) GRS data \citep{roman-duval2010}. While the latter is primarily used to derive gas velocity dispersion, we also use it to derive the morphology of the molecular clouds along with the line-of-sight mass contamination on our {\it Herschel}-based mass estimates (see Fig.~\ref{sketch}). For this purpose, we first need to compute $^{13}$CO-based H$_2$ column density cubes.

\subsubsection{$^{13}$CO column density cubes}

The Galactic Ring Survey $^{13}$CO(1-0) data have been used to compute H$_2$ column density cubes. To do this, we followed the exact same procedure as in \citet{roman-duval2010}. We give here a short description of this method. Towards each cloud of our sample, we compute the excitation temperature of the $^{13}$CO(1-0) line by using the $^{12}$CO(1-0) data from the UMSB survey and assuming that this line is optically thick. Then we make the further assumption that the excitation temperature of the $^{12}$CO(1-0) line is the same as that of the $^{13}$CO(1-0) line. With these assumptions, we can compute the $^{13}$CO(1-0) excitation temperature for every voxel. In some cases, where there are strong density/temperature  gradients not traced by $^{12}$CO(1-0)  (because it is optically thick) but traced by  $^{13}$CO(1-0) the estimated excitation temperature is not high enough. In such cases ($\sim 1\%$ of the voxels) we artificially increase the excitation temperature by 10\%, enough to get the excitation temperature larger than the $^{13}$CO(1-0) brightness temperature everywhere in our cloud sample. Combining the excitation temperature with the $^{13}$CO(1-0) cube one can compute the $^{13}$CO(1-0) opacity which is then converted into a $^{13}$CO column density, and finally into a H$_2$ column density assuming a constant $^{13}$CO abundance with respect to H$_2$ of $1.8\times10^{-6}$ for all clouds \citep{blake1987,langer1990}. This procedure provides us with cubes of H$_2$ column density for each IRDC. Note that some IRDCs are embedded within the same molecular clouds, and as a result we end up with 27 IRDCs embedded within 24 individual molecular clouds.

\subsubsection{Mass and velocity dispersion estimates}

For each molecular cloud we estimate two mass profiles, one using our {\it Herschel}-based column density images (see Sec.~3.1.1) and one using the $^{13}$CO-based column density cubes. Each of them are affected by different biases that can be, at least partially, removed by using the combination of both datasets. On one hand, far-infrared dust continuum emission of Galactic plane molecular clouds is mostly optically thin and traces the entire ISM, but suffers from line-of-sight confusion (see Fig.~\ref{sketch}). As a result, one cannot determine correctly the morphology of the clouds using our ${\it Herschel}$-based column density images and mass estimates are likely to be overestimated. On the other hand, $^{13}$CO-based column density cubes permit the identification of individual molecular clouds along the velocity axis, but can suffer from opacity effects and abundance variations. Because our mass estimates on clump scale are based on the {\it Herschel} column density images, we have decided to do the same for the diffuse parts of the clouds. However, we use the $^{13}$CO-based column density cubes to determine their morphologies, along with determining the percentage of multiple cloud line-of-sight mass contamination in our {\it Herschel}-based mass measurements. 

 Using our  $^{13}$CO-based column density cubes,  we first produced a 2D H$_2$ column density map by integrating the cube in a 5 to 10 km/s window centred on the cloud systemic velocity (see Fig.~\ref{example} and Appendix A). Then, we ran the same dendrogram analysis on this map as the one used on the clumps.  We then produced the average$^{13}$CO-based H$_2$ column density spectra for all identified groups of connected pixels. While the resulting spectra displayed in Fig.~\ref{example} for the SDC18.888-0.476 cloud show a relatively simple (even though non-Gaussian) single-peaked emission line, most clouds exhibit rather complex spectra often exhibiting multiple components with overlapping column density wings (see Appendix A). 

\begin{figure*}
   \centering
   \includegraphics[width=18.cm]{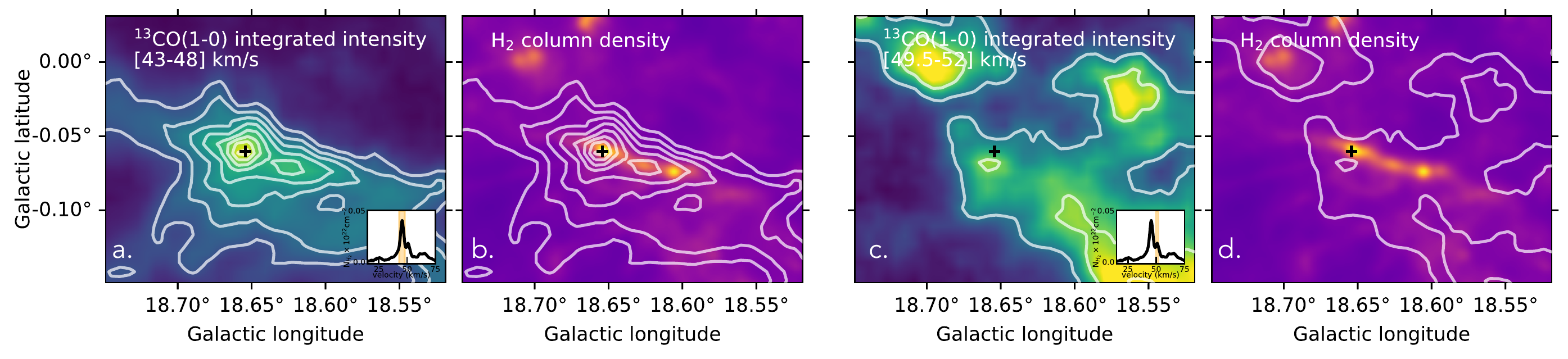}
   \vspace{-0.5cm}
      \caption{ (a) $^{13}$CO(1-0) integrated intensity map (colour and contours) of the lower velocity emission peak observed towards SDC18.624-0.070 at the native $44''$ GRS angular resolution. The plus sign shows the H$_2$ column density peak of the targeted IRDC. The average column density spectrum of the cloud is displayed as an inset in the bottom right corner, the orange shaded region shows the velocity integration interval. (b) {\it Herschel}-based H$_2$ column density map at $28''$ resolution of the same region and same contours as in (a). (c) and (d) show the same quantities as (a) and (b) but focussed on the high velocity emission peak observed towards SDC18.624-0.070.  }
         \label{sdc18p624_velcomp}
   \end{figure*}

With this in mind, we used multiple-Gaussian, up to a maximum of 4 components, to fit each spectrum using the python {\it curvefit} function (we also tested two other methods - see Appendix B). The mass $M_{\rm{^{13}CO}}$ and velocity dispersion $\sigma_{\rm{^{13}CO}}$ are then estimated using the following equations for the velocity dispersion:

\begin{equation}
\sigma_{\rm{^{13}CO}}=\sqrt{\sum_{i} w_i\left[(\varv_i-\bar{\varv})^2+\sigma_i^2\right]}
\label{vdisp13co}
\end{equation}

\noindent where the sum is over the Gaussian components, and $w_i$, $\varv_i$, and $\sigma_i$ are the weight, the central velocity and velocity dispersion of the $i^{\rm{th}}$ component, respectively. The centroid velocity $\bar{\varv}$ is obtained by:

\begin{equation}
\bar{\varv}=\sum_i w_i \varv_i
\end{equation}

\noindent And the weights are defined by:

\begin{equation}
w_i=\frac{m_i}{\sum_i m_i}
\end{equation}

\noindent where $m_i$ is the mass resulting from the integration of each individual Gaussian component, and:

\begin{equation}
M_{\rm{^{13}CO}}=\sum_i m_i
\end{equation}

\noindent The velocity dispersion calculated via Eq.~(\ref{vdisp13co}) includes two terms, i.e. the velocity dispersion from individual Gaussian components, along with the component-to-component centroid velocity dispersion. This is justified by the fact that we are here interested in estimating the entire kinetic energy budget of the clouds we are analysing. Note also that only the Gaussian components that we believe belong to the cloud of interest  are used for the determination of the mass and velocity dispersion. Those are identified by integrating, separately, each $^{13}$CO(1-0) emission peak and visually evaluate what peak best matches the morphology of the embedded IRDC. It is possible though that different components that we consider as being part of different molecular clouds are physically interacting with each other via, e. g., cloud-cloud collision. Such interactions can lead to the creation of intermediate velocity gas \citep{haworth2015, bisbas2017} for which it might become difficult to determine to which cloud it belongs, potentially leading to large uncertainties in the estimate of  $\sigma_{\rm{^{13}CO}}$.  In Fig.~\ref{sdc18p624_velcomp} we show the case of SDC18.624-0.070 for which it has been argued that such collision is currently occurring \citep{dewangan2018}. On that figure we display the $^{13}$CO(1-0) integrated intensity maps in the two velocity intervals that encompass the two emission peaks present in the region. Those maps are displayed at the native $44''$ resolution of the GRS survey. First we can see that the lower velocity component nicely matches the morphology of the {\it Herschel}-based H$_2$ column density obtained for that cloud, while the higher velocity component does not. This demonstrates that the gas traced by $^{13}$CO(1-0) is clearly associated to the targeted IRDC, and it also allows us to discard unrelated velocity components. However, in the case of SDC18.624-0.070 the morphology of the higher velocity cloud does indeed suggest an interaction with the lower velocity one. A similar exercise has been made for all clouds in order to ensure the correct association of the $^{13}$CO(1-0) components to each IRDC.

The radius of each dendrogram's connect group of pixels is given by:

\begin{equation}
R_{\rm{^{13}CO}}=\sqrt{\frac{n_{\rm{pix}}^{\rm{^{13}CO}}\Omega_{\rm{pix}}^{\rm{^{13}CO}}d^2}{\pi}}
\end{equation}

\noindent where $n_{\rm{pix}}^{\rm{^{13}CO}}$ is the number of pixels within each connected group of pixels and $\Omega_{\rm{pix}}^{\rm{^{13}CO}}$ is the solid angle subtended by a pixel. In parallel to these $^{13}$CO-based mass estimates, we derive {\it Herschel}-based ones. For this we use the exact same connected groups of pixels as those used above, but this time we use  our {\it Herschel}-based H$_2$ column density maps to obtain the masses via:

\begin{equation}
M_{Hers.}^{\rm{unc}}=\Omega_{\rm{pix}}^{\rm{^{13}CO}}d^2\mu_{\rm{mol}}m_{\rm{H}}\sum_{i=1}^{n_{\rm{pix}}^{\rm{^{13}CO}}}N_{\rm{H_2},i}
\end{equation}

\noindent where all parameters are identical to those presented in Eq.~(\ref{clumpmass}) and $M_{Hers.}^{\rm{unc}}$ stands for {\it uncorrected} {\it Herschel}-based masses. The reason why those are uncorrected is due to the contamination  of the mass estimates by the presence of multiple clouds along the line-of-sight. One can correct for this by estimating the fraction $f_{\rm{los}}$ of the total mass of molecular clouds along the line-of-sight that is locked up within the cloud of interest. That can be achieved by integrating the $^{13}$CO-based H$_2$ column density spectra across the entire GRS velocity range, along with integrating the best-fit Gaussian model for the cloud of interest. This can be formulated as:
\begin{equation}
f_{\rm{los}}=\frac{\int_{\rm{model}}N_{\rm{H_2}}^{\rm{^{13}CO}} d\varv}{\int_{\rm{all}}N_{\rm{H_2}}^{\rm{^{13}CO}} d\varv}
\end{equation}

\noindent This correction factor can be calculated for each dendrogram group of connected pixels and then be applied to the uncorrected masses via:
\begin{equation}
M_{Hers.}^{\rm{corr}}=f_{\rm{los}} M_{Hers.}^{\rm{unc}}
\end{equation}

\begin{figure}
  \vspace{-0.cm}
 \hspace{-.cm}
   \centering
   \includegraphics[width=8.5cm]{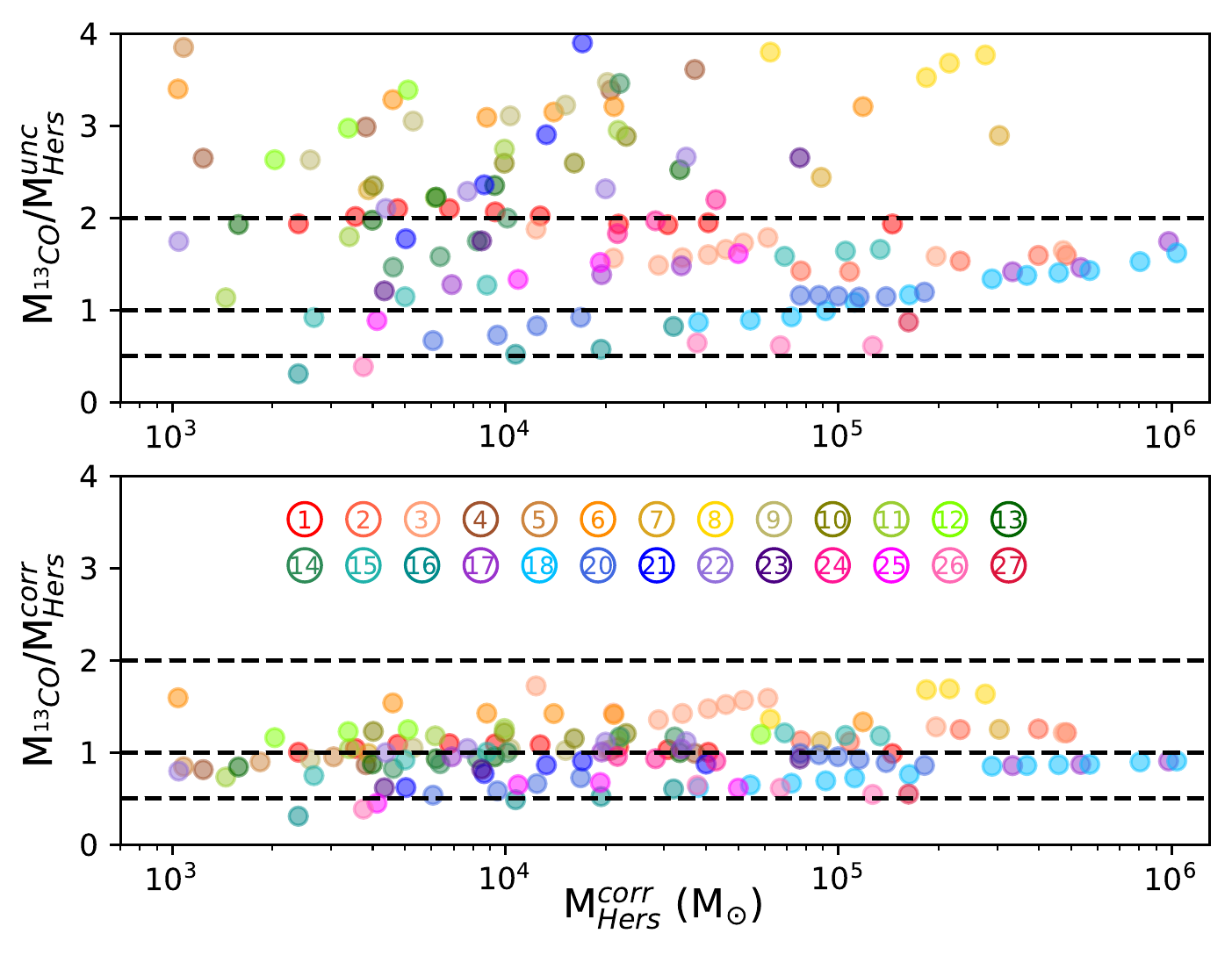} 
\vspace{-0.5cm}
      \caption{Comparison between the ratio of $^{13}$CO-based  and {\it Herschel}-based uncorrected/correct (top/bottom) cloud  masses for all clouds, at all radii, as a function of the corrected {\it Herschel}-based cloud  masses. }
         \label{diffmasscomp}
   \end{figure}

\noindent Figure~\ref{diffmasscomp} shows a comparison of the ratio of $M_{\rm{^{13}CO}}/M_{Hers.}^{\rm{unc}}$  versus $M_{Hers.}^{\rm{corr}}$ and  $M_{\rm{^{13}CO}}/M_{Hers.}^{\rm{corr}}$  versus $M_{Hers.}^{\rm{corr}}$. This figure clearly shows the vast improvement in the mass agreement once the correction factor is being applied. After correction, the masses agree within less than a factor 2 and we see little evidence for significant $^{13}$CO depletion, at least not on those scales. This excellent agreement also indicates that one can safely use the  $^{13}$CO(1-0) velocity dispersion measurements in conjunction with the {\it Herschel}-based  cloud masses. In the rest of this paper we will be using the {\it Herschel}-based corrected masses.

\subsection{Combined profiles}

 \begin{figure*}
  \vspace{-0.cm}
 \hspace{-.cm}
   \centering
   \includegraphics[width=18.cm]{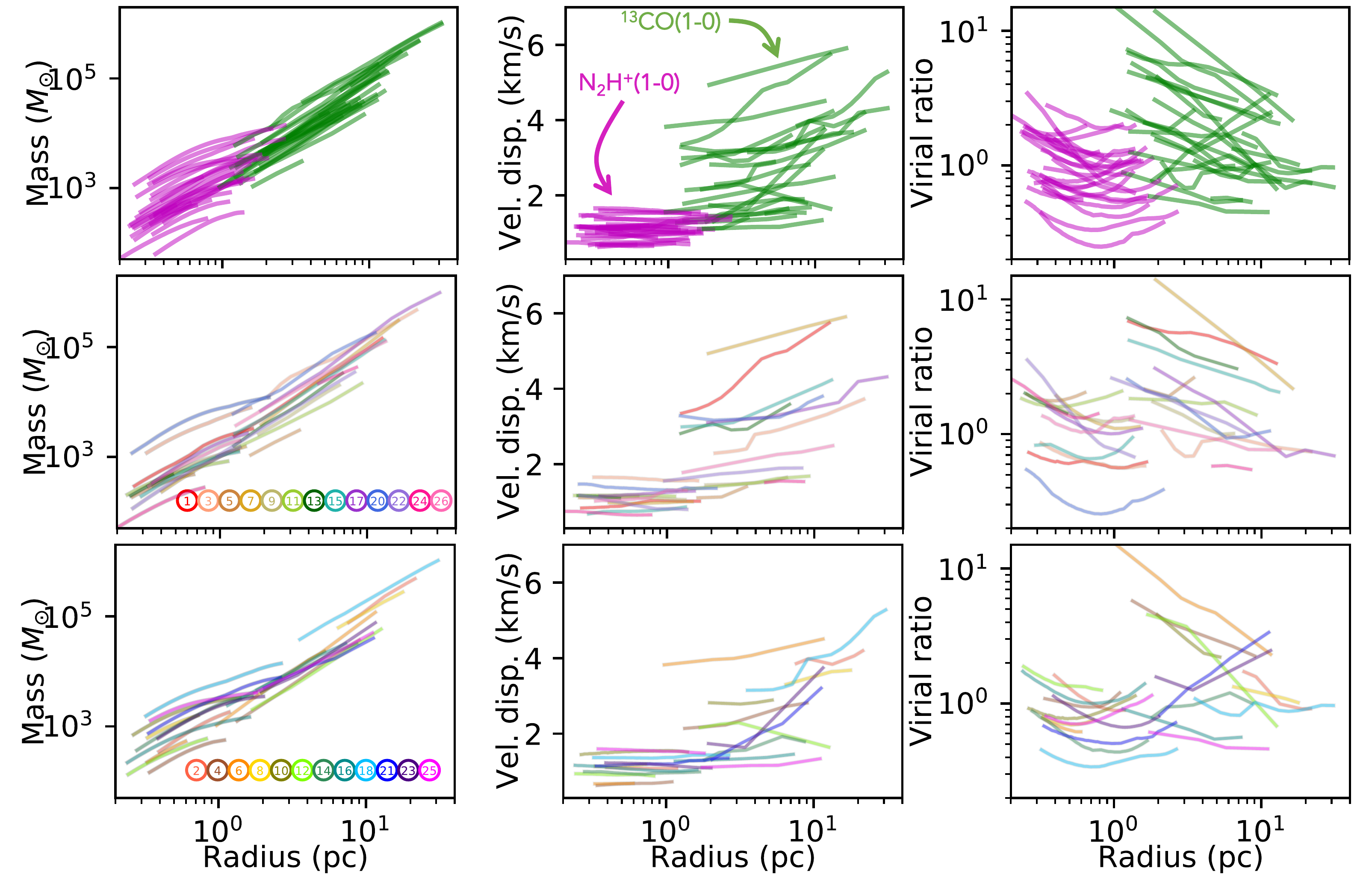} 
\vspace{-0.8cm}
      \caption{Profiles of all 26 infrared dark clouds and their parent molecular clouds from our sample. On the top row the purple points represent those for which the clump scale velocity dispersion has been measured using N$_2$H$^+$(1-0), and the green points are those for which the cloud scale velocity dispersion has been measured using  $^{13}$CO(1-0). The  middle and bottom rows show the same data point as the top row but each individual cloud/clump has a unique colour so that one can track their profiles. Half of the clouds have been plotted in each for clarity. (left): mass profiles $m(r)$; (middle): velocity dispersion profiles $\sigma_{\rm{tot}}(r)$; (right): virial ratio profiles $\alpha_{\rm{vir}}(r)$. }
         \label{obsprof}
   \end{figure*}

In this paper, we adopt a top-down approach by which, for every column density contours, we only analyse the one group of connected pixels that covers the position of the IRDC {\it Herschel}-based column density peak. As a result, sibling clumps that might be part of the same molecular clouds as our IRDC sample are not separately analysed, even though they contribute to the mass and velocity dispersion of the dendrogram structures that encompass both them and the targeted IRDC.

Figure~\ref{obsprof} shows the mass profiles $m(r)$ and the velocity dispersion profiles $\sigma_{\rm{tot}}(r)$ for the 26 clumps of our sample and their parent molecular clouds.  For the measurement on clump scales (the purple lines) we have $m(r)=M_{\rm{N_2H^+}}\left(R_{\rm{N_2H^+}}\right)$, while for the measurements on cloud scale (the green lines), we have $m(r)=M_{\rm{^{13}CO}}\left(R_{\rm{\rm{^{13}CO}}}\right)$. Also, the velocity dispersion $\sigma_{\rm{tot}}$ is the total (thermal+turbulent) line-of-sight velocity dispersion of the gas and is estimated using the observed velocity dispersion via:

\begin{equation}
\sigma_{\rm{tot}}^2=\sigma_{\rm{line}}^2+k_BT\left(\frac{1}{\mu_{\rm{mol}} m_{\rm{H}}}-\frac{1}{m_{\rm{mol}}}\right)
\end{equation}

\noindent where $\sigma_{\rm{line}}$ is the observed velocity dispersion of the gas as inferred from the observation of a given molecular line (N$_2$H$^+$(1-0)  for the purple points, and  $^{13}$CO(1-0)  for the green points), $T$ is the gas temperature, $m_{\rm{mol}}$ is the mass of the observed molecule (here $m_{\rm{mol}}= m_{\rm{N_2H+}}=m_{\rm{^{13}CO}}=29m_{\rm{H}})$, $\mu_{\rm{mol}}$ is the molecular weight which is here taken to be 2.33, and $m_{\rm{H}}$ is the mass of the hydrogen atom. We here assume a gas temperature of 15~K for all clouds, which is the average temperature measured within infrared dark clouds \citep[e.g.][]{peretto2010b,battersby2011}. The impact of that assumption is negligible for most velocity dispersion measurements, and only have a measurable impact for velocity dispersions $\le1$~km/s. Finally, the right-hand-side panel of  Fig.~\ref{obsprof} shows the corresponding viral ratio profiles  $\alpha_{\rm{vir}}(r)$. Virial ratios, defined as $\alpha_{\rm{vir}}=2E_{\rm{K}}/|E_{\rm{G}}|$ with $E_K$ the kinetic energy of the gas and $E_{\rm{G}}$ its gravitational energy, provide a zero-order measure of a cloud dynamical state. While the kinetic energy of a cloud can be relatively easily estimated, the estimate of its gravitational energy usually requires to make simplifying assumptions on the morphology and density profile of the cloud. \citet{bertoldi1992} have evaluated $E_{\rm{G}}$ in case of different power law densities and different cloud aspect ratios.  They show that for cloud with aspect ratios lower than 10 (as it is the case in this study) $|E_{\rm{G}}|$ is only decreased by a maximum of 8\% compared to the spherical case. However, for clouds that have power law density such as $\rho\propto r^{-\gamma}$ with $\gamma=2$, $|E_{\rm{G}}|$ is increased by 67\%. The impact of the density gradient on $|E_{\rm{G}}|$ is stronger than the non-sphericity of the cloud. For simplicity, most studies of the virial ratio of molecular clouds usually approximate them as uniform density spheres, which is also what we will do, and discuss correction factors later. In this case, one can show that the virial ratio $\alpha_{\rm{vir}}(r)$ is given by:

\begin{equation}
\alpha_{\rm{vir}}(r)=5\frac{\sigma_{\rm{tot}}^2 r }{Gm}
\end{equation}

\noindent Figure~\ref{obsprof} shows a number of important features. In the following we will discuss those separately.

\subsubsection{Observed mass profiles}

The mass profiles presented in Fig.~\ref{obsprof}  spread over  4 orders of magnitude in mass and 2 orders of magnitude in radius. The masses estimated on cloud scale (the green lines) and clump scale (the purple lines) mostly connect at a scale of about 2\,pc, which corresponds to the maximum extent of the N$_2$H$^+$(1-0) emission and half the resolution ($3'$) of the $^{13}$CO-based column density images that we use to derive the morphology of the clouds. Note that, even though derived from the same {\it Herschel}-based H$_2$ column density maps, the masses on clump and cloud scales do not produce continuous mass profiles. The reason for this is that we are removing the column density background $N_{\rm{N_2H^+}}^{\rm{edge}}$ to every clump scale mass measurements so that the velocity dispersion estimate is that of the measured gas mass.  Finally, when looking at shapes of the profiles, we notice that the clump scale mass profiles are more curvy and and exhibit shallower gradients than those from the more diffuse parts. 

\subsubsection{Observed velocity dispersion profiles}

The velocity dispersion profiles presented in Fig.~\ref{obsprof}  are the most striking. First, there is a clear discontinuity between the velocity dispersion measurements obtained on clump scale and those obtained on cloud scale. Having such different measurements clearly indicates that there is a systematic bias in the method that is being used to perform those measurements. Second, the shapes of the profiles are also strikingly different. While on the largest scale, the velocity dispersion mostly decreases with decreasing radius, on the smallest scale, the velocity dispersion profiles are mostly flat. This is very different from a typical Larson-type relation \citep{larson1981} for which we would expect the velocity dispersion to decrease down to the sonic-scale at about 0.1\,pc.

The method used to derive the velocity dispersions of the often complex $^{13}$CO(1-0) spectra may have an impact of the observed discontinuity. As presented in Appendix B, in addition to the multiple Gaussian fitting, we also applied two other methods, i.e. a standard moment method, along side what we call the Peak method. The latter is based on the FWHM of the main emission peak and tends to exclude low-intensity high-velocity wings from the velocity dispersion measurements. While the moments method increases even further the discontinuity between clump and cloud scales, the Peak method slightly decreases it, with a larger fraction of cloud exhibiting relatively flat velocity dispersion profiles all the way up to tens of parsecs (see Fig.~ B3). However, the overall behaviour of the profiles remain very similar to what is obtained when using the Gaussian fitting method.

\subsubsection{Observed virial ratio profiles}

Since the virial ratio profiles presented in Fig.~\ref{obsprof} are built from the mass and velocity profiles, they carry similar features. For instance, the virial ratios present a discontinuity at the around $r=2$\,pc, which is the consequence of the discontinuity observed in the velocity dispersion profiles. Note, however, that this discontinuity is attenuated as a result of the slightly larger masses estimated from the cloud scale measurements at that radius. Also, it is pretty clear that for most of the clouds, the virial ratios on the large scales (green lines) increase as the radius decreases. This trend has already been observed by \citet{hernandez2015} who interpreted it as a sign of CO depletion. Finally, the virial ratios estimated on clump scales (purple lines) present a curvy shape, which is the direct reflection of the curvy mass profiles observed on the same scales.  

\begin{figure*}
  \vspace{-0.cm}
 \hspace{-.cm}
   \centering
   \includegraphics[width=18.cm]{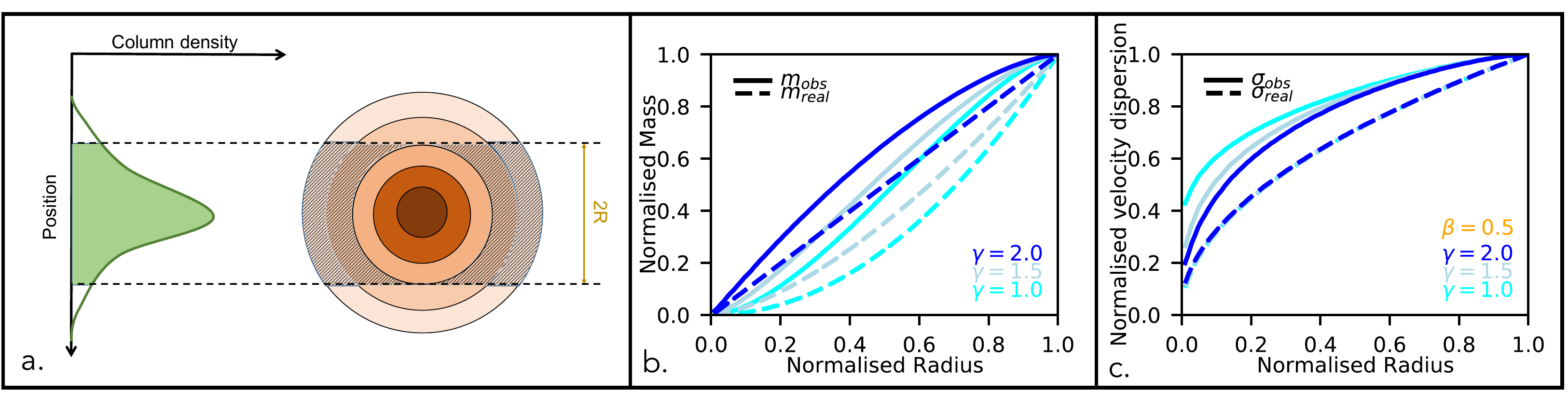} 
\vspace{-0.5cm}
      \caption{(a) Sketch illustrating measurement biases once a 3D cloud is being projected onto the plane of sky. In this case,  a spherical cloud (on the right) whose volume density increases towards the centre has a projected column density profile that is represented by the plot on the left. The bjjective mass estimate within projected radius $R$ will be that represented by the green shaded area, which includes material  that is not part of the volume of the sphere of radius $R$ (region of the cloud that is barred). (b): Normalised mass profiles of a spherical cloud with 3 different density profile $\rho\propto r^{-\gamma}$, with $\gamma=(2.0,1.5,1.0)$. The solid lines show the masses as observed, while the dashed lines show the real mass enclosed within a given radius. (c): Normalised velocity dispersion profiles of a spherical with the same density profiles as in (b) and with a velocity dispersion profile $\sigma\propto r^{\beta}$ with $\beta=0.5$.}
         \label{sketch2}
   \end{figure*}

\subsection{Uncertainties}

There are a number of uncertainties that we need to consider when interpreting the profiles presented in Fig.~\ref{obsprof}. First, there are uncertainties that do not affect the shape of the profiles but do impact their overall scaling. One example of such uncertainty is the distance to the clouds which is typically 10\% to 20\% \citep{reid2009}. If our IRDC-hosting cloud are not located at the near distance though then the distance could be 4 times larger for some clouds (see Fig.~\ref{distcomp}). That uncertainty will impact the mass and radius measurements uniformly across the profile of an individual cloud. Second, there are uncertainties that can potentially impact the shape of individual profiles. Regarding the mass profiles, the assumption of a single temperature along the line-of-sight could potentially have an impact on the shape of the observed profiles. However, as we have shown in Sec. 3.1.3, the impact on the shape of the profile is minimal, while the impact on the absolute mass values can be impacted by $20\%$ on average. Another uncertainty is related to the dust emissivity, i.e. $\kappa_{\lambda}$, we used when computing the {\it Herschel}-based H$_2$ column density maps. In this study we used the same dust emissivity law for the clump scale  and cloud scale measurements. It is however well known that dust emissivity changes with density and temperature \citep[e.g.][]{ysard2015,sadavoy2016}.  At this point we have no means to set strong constraints on this particular aspect of dust property uncertainties, but the law we adopted has been shown to be compatible with dust emission in both the more diffuse \citep{planck2011} and denser \citep{rigby2018} gas environments. Also, as it can be seen in Fig.~\ref{diffmasscomp}, the {\it Herschel}-based masses are within a factor of two of the $^{13}$CO-based masses which use a completely different set of assumptions. This suggests that, if dust properties do change across the radial profiles of molecular clouds, this does not have a dramatic effect on our mass estimates. Finally, the uncertainty related to our choice of  $N_{\rm{N_2H^+}}^{\rm{edge}}$ (see Sec.~3.1.3) has a direct impact on the clump scale mass estimates with a $\sim10\%$ to $\sim30\%$ uncertainty for most clumps (see Table~2). This fractional mass uncertainty is not constant across the clump radial profiles and therefore can affect the mass profile shape. However, after computing the clump mass profiles with a representative range of  $N_{\rm{N_2H^+}}^{\rm{edge}}$ we can confirm that their overall shapes are barely affected (see Appendix C for the special case $N_{\rm{N_2H^+}}^{\rm{edge}}=0$).

Regarding uncertainties on the velocity dispersion, the N$_2$H$^+$(1-0) and $^{13}$CO(1-0) measurements differ. Indeed, the N$_2$H$^+$(1-0) velocity dispersion measurements are very well constrained, and have uncertainties that are of the order of $\sim0.1$~km/s. This implies that the flat velocity dispersion profiles observed on clump scale are very robust. Uncertainties on the $^{13}$CO(1-0) velocity dispersion measurements are a lot more variable from cloud-to-cloud depending on how complex the $^{13}$CO(1-0) spectra are. For the simple cases, such as SDC18.888-0.476 (see Fig.~\ref{example}) the uncertainty is of the order of $\sim0.2$~km/s, however is can be as high as $\sim1$ km/s in more complex cases such as SDC18.624-0.070 (see Fig.~A1). These larger uncertainties are also reflected by the large differences in velocity dispersion measurements when using different evaluation methods (see Fig.~\ref{compmeth}).

Overall, while the inherent uncertainties on the different quantities presented in Fig.~\ref{obsprof} might shift the profiles up and down, their shapes are fairly robust and are likely to be a true representation of how the projected mass, velocity dispersion, and virial ratio profiles of clumps and clouds behave.

\section{Spherical models}

As discussed above, the profiles displayed in Fig.~\ref{obsprof} present a number of characteristic features. Before interpreting them one needs to be aware a few biases that exist and that we may be able to quantify. First, masses, as presented in the left-hand-side panel of Fig.~\ref{obsprof}, have been computed using the bijective mass estimates \citep{rosolowsky2008}. Such masses are always overestimated as a consequence of cloud material lying along the line-of-sight which is not part of the closed volume of radius $r$ (see Fig.~\ref{sketch2}). The impact of using the bijective method to estimate masses at different radius is illustrated in Fig.~\ref{sketch2}b. Second, the velocity dispersion measurements are being done on spectra that also include the same unrelated line-of-sight material which may have larger or smaller velocity dispersion than the gas lying within the volume of interest. Depending on the exact shape of the combined density and velocity dispersion profiles, this might lead to over or underestimated  observed velocity dispersions (see Fig.~\ref{sketch2}c). Also, the clump and cloud scale measurements are derived at different angular resolutions, and a background column density has been subtracted to the former and not to the latter. The impact of all those on the observed profiles is unclear. The purpose of the models presented in the rest of this section is to quantify the impact of projection on the mass and velocity dispersion measurements in relation to the observed profiles \citep[for a similar approach on core scale see][]{singh2021}. We do not attempt to fit the profiles of individual clouds as spherical clouds are not a good representation of the complex density structures of molecular clouds.

 \begin{figure*}
  \vspace{-0.cm}
 \hspace{-.cm}
   \centering
   \includegraphics[width=18.cm]{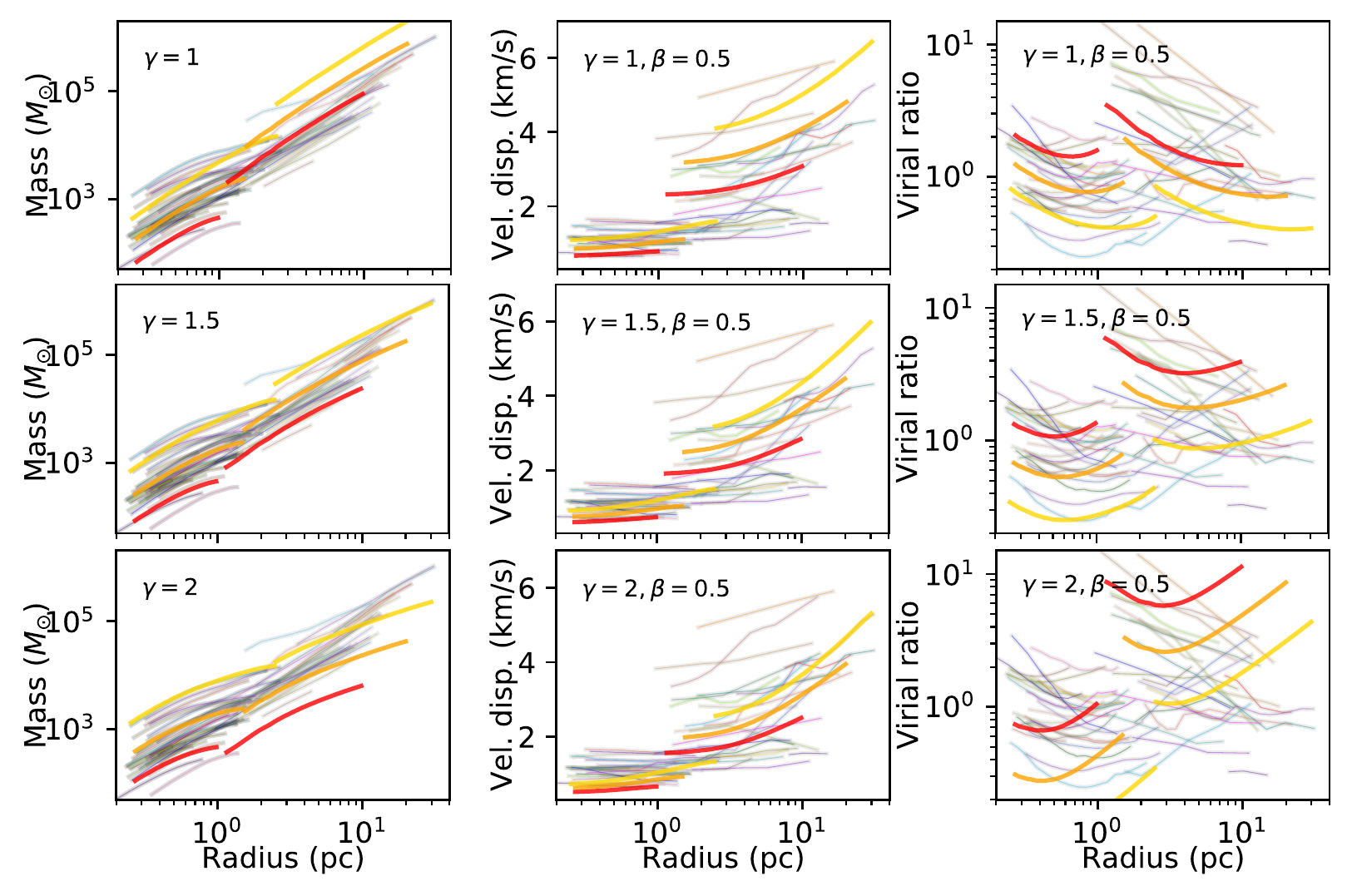} 
\vspace{-0.8cm}
      \caption{Mass, velocity dispersion, and virial ratio profiles (from left to right), for three different $\gamma$ values (from top to bottom: $\gamma=[1.0,1.5,2.0]$). The grey lines are the same observed data points as presented in Fig.~\ref{obsprof}. The red, orange, and gold lines are three different spherical models with three different normalisations such that  they cover the range of masses and velocity dispersions as measured on parsec scale (i.e. on scales representative of $R_{\rm{N_2H^+}}^{\rm{start}}$ in Table 2). All models have the same velocity dispersion profile exponent, i.e. $\beta=0.5$.}
         \label{model_prof}
   \end{figure*}

\subsection{Single power-law profiles}

We first consider models with single density and velocity dispersion power-law such as:

\begin{equation}
\rho(r)=\rho_0\left(\frac{r}{r_0}\right)^{-\gamma}
\end{equation}

\begin{equation}
\sigma(r)=\sigma_0\left(\frac{r}{r_0}\right)^{\beta}
\end{equation}

\noindent where $\rho_0$, $r_0$ and $\sigma_0$ are normalisation constants. For a given pair of $\gamma$ and $\beta$ values, we numerically construct a spherical cloud of a given mass and radius that we then project on the plane-of-the-sky in order to construct mass surface density maps (see Appendix D for more details). We do this last operation twice, once up to radius $R_{\rm{end}}^{\rm{^{13}CO}}$ and once up to radius $R_{\rm{end}}^{\rm{N_2H^+}}$,  $R_{\rm{end}}^{\rm{^{13}CO}}$ and $R_{\rm{end}}^{\rm{N_2H^+}}$ being the radii at which $^{13}\rm{CO}$(1-0) and $\rm{N_2H^+}$(1-0) emission becomes undetectable.\footnote{Note that the $R_{\rm{end}}$ parameters are defined pre-convolution and as such  do not exactly match the $R_{\rm{edge}}$ parameters defined in Sec.~3 that are obtained directly from the observations (i.e. post-convolution).} We then convolve each mass surface density images at the resolution of our observations. Finally, we integrate both mass surface density images at various radii to derive their projected mass profiles.
Regarding the velocity dispersion profile, we first weight the velocity dispersion at each radius, in 3D,  by the local mass density. We then project this quantity onto the plane-of-the-sky, and then integrate the resulting maps at various radii. Finally, we divide these profiles by the corresponding mass profiles in order to obtain projected mass-weighted velocity dispersion profiles.

 \begin{figure*}
  \vspace{-0.cm}
 \hspace{-.cm}
   \centering
   \includegraphics[width=18.cm]{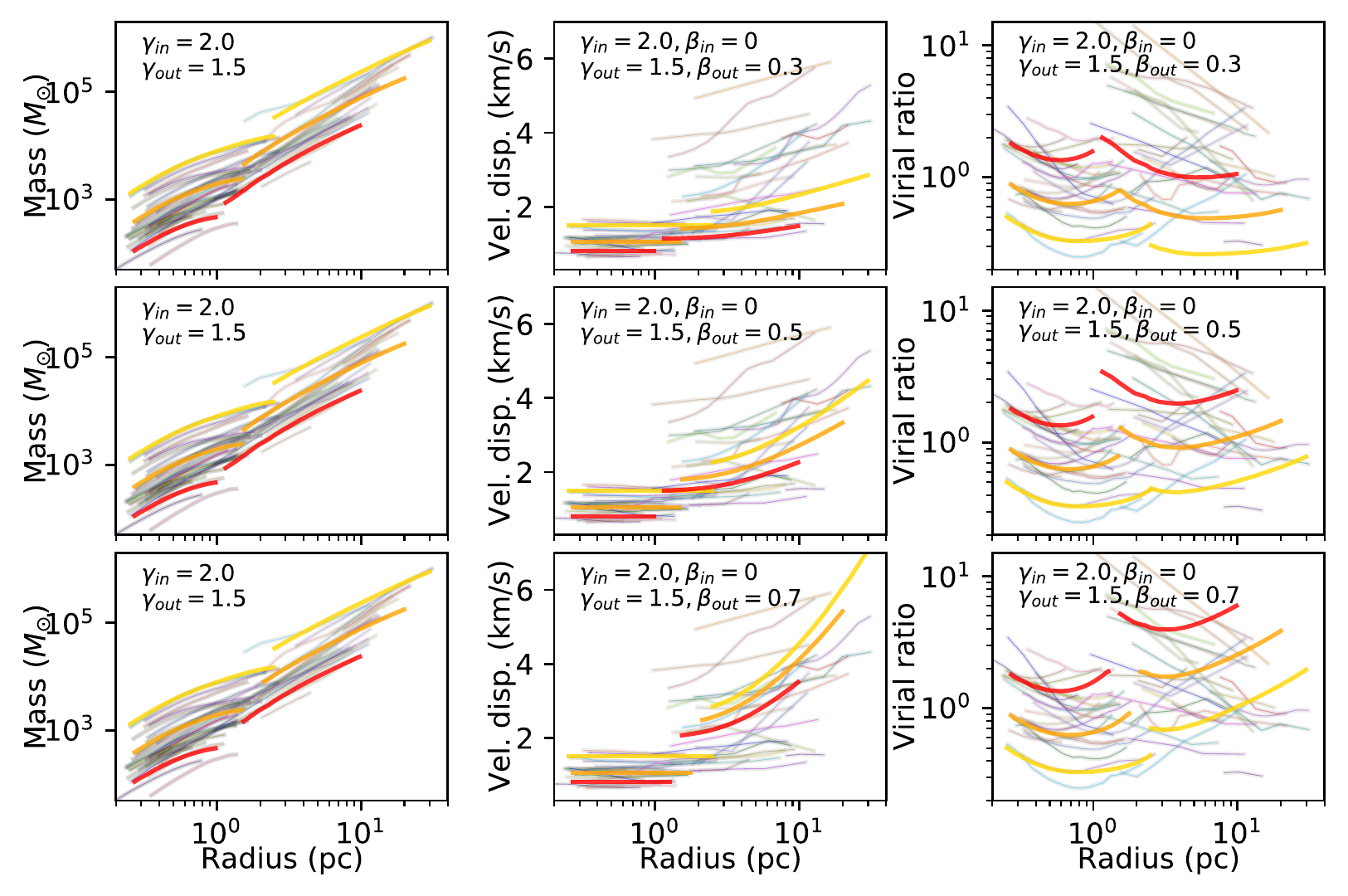} 
\vspace{-0.8cm}
      \caption{Same as Fig.~\ref{model_prof} but for broken power-law profiles. Each row now corresponds to a different $\beta_{\rm{out}}$ value  (from top to bottom: $\beta_{\rm{out}}=[0.3,0.5,0.7]$). For all models, $\beta_{\rm{in}}$, $\gamma_{\rm{in}}$, and $\gamma_{\rm{out}}$ are fixed to [0.0, 2.0,1.5], respectively. }
         \label{model_prof_broken}
   \end{figure*}

 \begin{figure*}
  \vspace{-0.cm}
 \hspace{-.cm}
   \centering
   \includegraphics[width=18.cm]{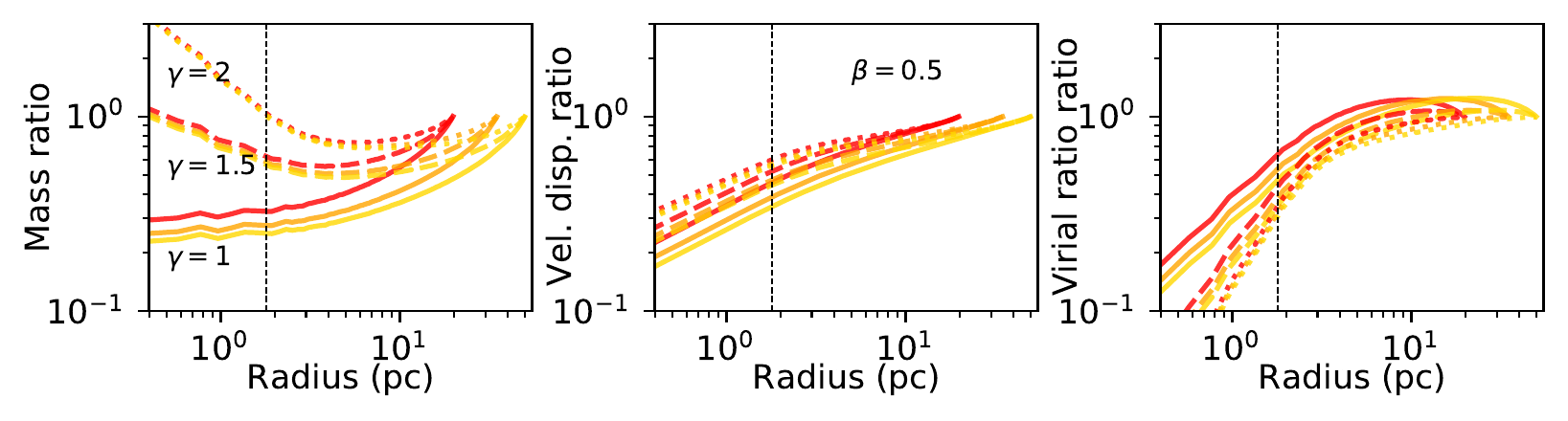} 
\vspace{-0.8cm}
      \caption{Ratios of the true over projected mass, velocity dispersion, and virial ratio profiles for all 9 models displayed in Fig.~\ref{model_prof}. The projected models are at the same angular resolution as our $^{13}$CO-based column density maps, that is $3'$. The solid, dashed, and dotted profiles correspond to the density profile indices $\gamma=[1,1.5,2]$, respectively. All models have the same velocity dispersion profile index $\beta=0.5$. The three colours correspond to different normalisations. The vertical black dashed line shows half of the $3'$ beam size at a distance of 4~kpc.}
         \label{prof_ratios}
   \end{figure*}

In the models presented here, there are  essentially four free parameters, i.e $\gamma$, $R_{\rm{end}}^{\rm{N_2H^+}}$ ($=r_0$), $R_{\rm{end}}^{\rm{^{13}CO}}$ , and $M_{\rm{end}}^{\rm{N_2H^+}}$, for the mass profiles, and an additional two free parameters, i.e. $\beta$, and  $\sigma_{\rm{end}}^{\rm{N_2H^+}}$ ($=\sigma_0$), for the velocity dispersion profiles. The parameter $\rho_0$ is derived from $\gamma$, $R_{\rm{end}}^{\rm{N_2H^+}}$, and $M_{\rm{end}}^{\rm{N_2H^+}}$ and is, thus, not a free-parameter of the models. As already mentioned, the purpose of those models are not to find a set of best parameters for each individual clouds, but rather to understand the trends that are present in the cloud sample. With that in mind, Fig.~\ref{model_prof} shows a set of 9 models against  the observed profiles. The normalisation of those models is such they match the range of mass and velocity dispersion at parsec-scales. Each row corresponds to a different $\gamma$ value but the same $\beta$ value. In each row, the three panels correspond to the mass, velocity dispersion, and virial ratio profiles. There are a number of important features in those models that we can notice straight away.  First, regarding the mass profiles, one can see that the cases $\gamma=1$ and $\gamma=2$ over-predict and under-predict, respectively, the mass of the clouds on the largest scales. We also notice that, while the $\gamma=1.5$ case provides a better overall agreement with the observed profiles, the profile shapes provided by the cases $\gamma=2$ and $\gamma=1$ seem to give a better match to the inner and outer parts, respectively,  of the observed profiles. We also notice that we successfully reproduce the curved shape of the inner parts of the profiles. 

Moving on to the velocity dispersion profiles displayed in Fig.~\ref{model_prof} it is clear that the simple 1D models represented here manage to reproduce the velocity dispersion discontinuity, in particular for the $\gamma=1$ case. The reason for this is that, for that series of models, a larger fraction of the mass is at low density where the velocity dispersion is the largest and as a result of the projection, the mass weighted velocity dispersion is overestimated by a large factor, up $\sim3$ in the $\gamma=1$ case. Also, while the $\gamma=1$ case manages to reproduce in a satisfactory way the shape and amplitude of the outer velocity dispersion profile, it cannot entirely match the observed flat velocity dispersion profiles in the inner regions. And this gets worse when considering steeper density profiles, i.e. $\gamma=1.5$ and $\gamma=2$.

Finally, looking at the virial ratio profiles in the last column of Fig.~\ref{model_prof}, we notice that none of them are completely satisfactory when compared to the observed profiles, even though one could argue that the shallower density models do better than the steeper ones. 

From this comparison between single-power law models and observed profiles, it seems clear that most of the observed features can be reproduced, at least to some extent, providing strong evidence that projection biases are mostly responsible for them. This comparison also shows that the single power-law models are limited and do not allow us to reproduce both the the inner and outer parts of the observed mass  and velocity dispersion profiles. Most noticeable is the velocity dispersion profiles for which the flat inner profiles are clearly different from the outer profile shapes.  

\begin{figure}
  \vspace{-0.cm}
 \hspace{-.cm}
   \centering
   \includegraphics[width=9.5cm]{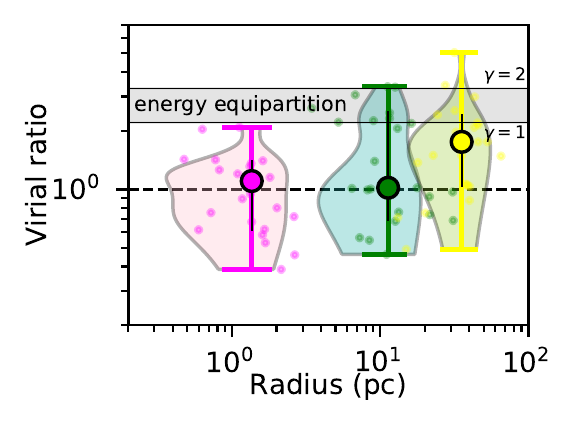} 
\vspace{-0.8cm}
      \caption{Violin plots of the virial ratios obtained on the largest scales in both N$_2$H$^+$(1-0) (magenta) and $^{13}$CO(1-0) (green), along with that obtained by \citet{miville-deschenes2017} in $^{12}$CO(1-0) (yellow) for the same sample of clouds. Each violin plot is located, along the x-axis, at the median radius value of each group. The median virial ratio value for each group is represented by a coloured circular symbol with a black edge, while the 16$^{\rm{th}}$ and 84$^{\rm{th}}$  percentile ranges are represented by vertical solid black lines. We have also overplotted the corresponding individual measurements as coloured circular symbols. The horizontal black dashed lines show virial ratio values $\alpha_{\rm{vir}}=1$ while the shaded area show the region of energy equipartition for density profile indices between $\gamma=1$ and $\gamma=2$. }
         \label{vir_comp}
   \end{figure}

\subsection{Broken power-law profiles}

In this section, we extend the models presented above from single power-law profiles to broken power-law profiles. More explicitly, we set two power-law exponents for both the density and velocity dispersion profiles defined as:

\begin{equation}
\gamma_{\rm{in}}  \hspace{0.2cm} {\rm and} \hspace{0.2cm} \beta_{\rm{in}}  \hspace{0.2cm} {\rm for} \hspace{0.2cm} r< r_0 \\\\
\gamma_{\rm{out}}  \hspace{0.2cm} {\rm and} \hspace{0.2cm} \beta_{\rm{out}}  \hspace{0.2cm} {\rm for} \hspace{0.2cm} r> r_0
\end{equation}

\noindent The method used to create the profiles is identical to that presented in the previous section. Based on our single power-law models, it seems  that the inner density profiles are, on average, steeper than the outer ones. Therefore, in the series of models presented in Fig.~\ref{model_prof_broken}  we used $\gamma_{\rm{in}}=2$ and $\gamma_{\rm{out}}=1.5$. Regarding the velocity dispersion profiles it seems clear that the velocity dispersion on clump scale is rather flat, with apparent very little variation in the profiles. On the other hand, the velocity dispersion profiles on cloud scale are diverse, both in terms of shape and normalisation. Therefore, each row in Fig.~\ref{model_prof_broken} corresponds to a different value of $\beta_{\rm{out}}$, while $\beta_{\rm{in}}$ is fixed to 0 for all models. In that figure one can see that we now reproduce rather well the average shape and magnitude of the mass profiles, and similarly for the velocity dispersion profiles on clump scales. One can also see that we do reproduce well some of the velocity dispersion profiles on cloud scale, although we fail in reproducing the low-mass high velocity dispersion profiles that populate the top part of the velocity and virial ratio profiles.   This is where our simple 1D models reach their limitations. Indeed, if one looks at the large scale mass distribution of those clouds (via the $^{13}$CO-based H$_2$ column density maps - see Fig.~\ref{example} and Appendix A), one can see that the clump we are focussing on does not dominate the mass on those scales, with many or more sibling clumps being present in the same parent cloud. As a result, the large scale velocity dispersion measured towards the  clumps of interest is mot likely driven by the presence of its siblings.  This cannot be reproduced with spherical models. Nevertheless, what this is showing is that lower gas density layers of high velocity dispersion gas surround those clumps, generating steep velocity dispersion discontinuities in their profiles.

\subsection{Projected versus true profiles}

 In the previous sub-sections we have characterised the origin of observed profile features with the help of projected models. We have not yet evaluated how the profiles of those same projected models compare to their own input profiles. Such comparison can be useful when it comes to understanding the impact of different aspects of the projection process has on the observed absolute profile values. 

Figure \ref{prof_ratios} shows the ratio of the true mass, velocity dispersion, and virial ratio profiles over those observed at $3'$ resolution, for all single power-law models presented in Figs.~\ref{model_prof}. We focus on the  cloud-scale profiles since the conclusions are the same for the clump-scale part of the models. One can see that, with the exception of the mass profile $\gamma=2$ (see below), the observed values are nearly systematically overestimated, whether it is mass, velocity dispersion, and virial ratios. However, regarding the virial ratio profiles, one can see that the observed values are within 20\% of the true value except for the inner radii, when one gets within a couple of angular resolution elements.

 Often in the literature one finds that radii of structures are being deconvolved from the beam size of the telescope. This is a valid approach, particularly for point-like sources as the measured fluxes come from a region of the sky that is necessarily smaller than what is observed after beam convolution. However, when dealing with molecular cloud measurements, the picture is not that clear as, at least in our case, two competing effects come into play: beam convolution and projection effects. While for a centrally concentrated density profile beam convolution will tend to spread the flux and mass to larger radii, the  line-of-sight integration of flux that does not come from within the volume of interest will tend to increase the flux/mass at a given projected radius. In Fig.~\ref{prof_ratios} (left) one can see that the relative impact of both effects depends on the density profile index and radius, convolution having the strongest impact for $\gamma=2$ at small radii resulting in mass underestimation, while projection tends to overestimate masses for any other combination of radius and $\gamma$. It thus becomes clear that projection is the dominant factor in terms of mass estimates accuracy.

 We have to note here that the direct comparison of our observed profiles with the modelled ones assumes that the tracers we use (i.e. dust continuum, $^{13}$CO(1-0) and N$_2$H$^+$(1-0) emission) reliably trace the energetics of the underlying clouds. In order to show that this is the case, radiative transfer calculations of our 1D models would have to be made. While this is deferred to a future paper that will look into a larger range of tracers, we argue that the large number of similarities  between observed and  modelled profiles is already evidence that the combination of tracers we use here is good enough for the purpose they serve.

\section{Discussion}

\subsection{Self-gravitating molecular clouds}

The question of the gravitational binding of molecular clouds has been, and still is, the subject of numerous debates \citep[e.g.][]{heyer2009, dobbs2011, ballesteros2011, miville-deschenes2017, vazquez-semadeni2019}. Here, we have all the necessary information to check whether the clouds we selected are  consistent with being gravitationally bound or whether there is a scale at which they switch from being bound to unbound. The profiles displayed in Fig.~\ref{obsprof} show that observed virial ratios for our cloud sample are nearly systematically below $\alpha_{\rm{vir}}=3$ at all radii. Most of the exceptions correspond to the measurements made at the smallest radii of the $^{13}$CO(1-0)-based profiles. As our models showed (see Fig.~\ref{model_prof}) increased virial ratios with decreasing radii can be reproduced when large layers of high velocity dispersion gas lay along the line-of-sight and contaminates the measurements. As a result the most reliable measurements are those obtained on the largest scales (see Fig.~\ref{prof_ratios}). Figure~\ref{vir_comp} shows the distributions of virial ratios obtained at those largest scales for both N$_2$H$^+$(1-0) and $^{13}$CO(1-0) measurements. 

For uniform density spheres, the transition between gravitationally bound and unbound gas occurs at $\alpha_{\rm{vir}}=2$, while for clouds with density profiles such as $\rho\propto r^{-1}$, $\rho\propto r^{-1.5}$, and $\rho\propto r^{-2}$, the limit moves up to 2.2, 2.5, and 3.3, respectively.  Correction factors regarding the non-spherical shape of clouds are less than 8\% as long as the aspect ratio of the clouds is lower than 10 \citep{bertoldi1992}, which is the case for all clouds in the sample. For non-uniform velocity dispersion profiles correction factors also exist \citep{miville-deschenes2017}, but these are of the order of 5\% for the diffuse parts of the cloud and non-existent for the dense part (see Appendix E). Figure~\ref{vir_comp} reveals that $\sim85\%$ of $^{13}$CO-based measurements, and 100$\%$ of the N$_2$H$^+$-based measurements have $\alpha_{\rm{vir}}\le2.5$. 

 Virial ratios as estimated here only include the kinetic and gravitational energy volume terms of the virial theorem. However, the surface terms can also be important to consider when evaluating whether a piece of molecular gas is gravitationally bound or not. \citet{dib2007}  showed that,  in the context of magneto-hydrodynamical simulations of turbulent molecular clouds, the surface kinetic energy of cores can be as large as its volume counterpart and be responsible for tearing them apart despite having, sometimes, virial ratios consistent with being self-gravitating. By applying a similar approach to their own simulations, \citet{weiss2022} also showed that  clumps' surface terms can be significant and even govern their dynamical evolution, although clumps that are dense and massive enough to form cores have dominant volume terms and mostly self-gravitating virial ratios. Measuring the surface terms of observed clumps is, in practice, impossible since one only gets to measure cloud properties once projected onto the plane-of-the-sky. However, by measuring virial ratios at different radii within clouds we ensure that the surface kinetic energy at a given radius becomes part of the volume kinetic energy at larger radii. As a result, virial ratio profiles of unbound clouds with large surface kinetics energy terms should exhibit self-gravitating virial ratios only at a minority of radial points. Hence, the fact that the observed virial ratios of our clump and molecular cloud sample are consistently below 3 across a large range of spatial scales (see Fig.~\ref{vir_comp}) strongly supports the idea that the vast majority of the molecular clouds, if not all, is self-gravitating from tenths of a parsec up to several tens of parsecs.

\begin{figure}
  \vspace{-0.cm}
 \hspace{-.cm}
   \centering
   \includegraphics[width=8.5cm]{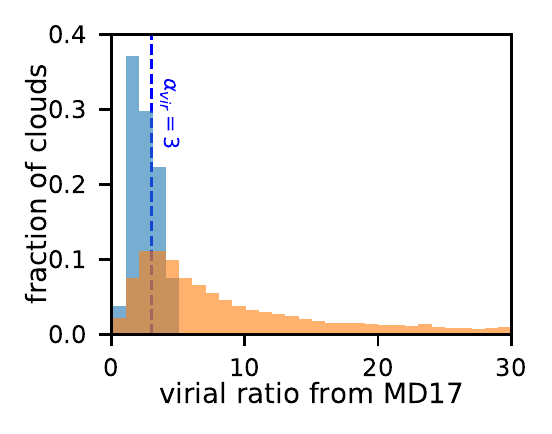} 
\vspace{-0.5cm}
      \caption{Distributions of virial ratios as estimated by \citet{miville-deschenes2017} for the 24 clouds presented here (blue histogram) and their entire cloud population (orange histogram).}
         \label{avirMD17}
   \end{figure}

\begin{figure*}
  \vspace{-0.cm}
 \hspace{-.cm}
   \centering
   \includegraphics[width=18.cm]{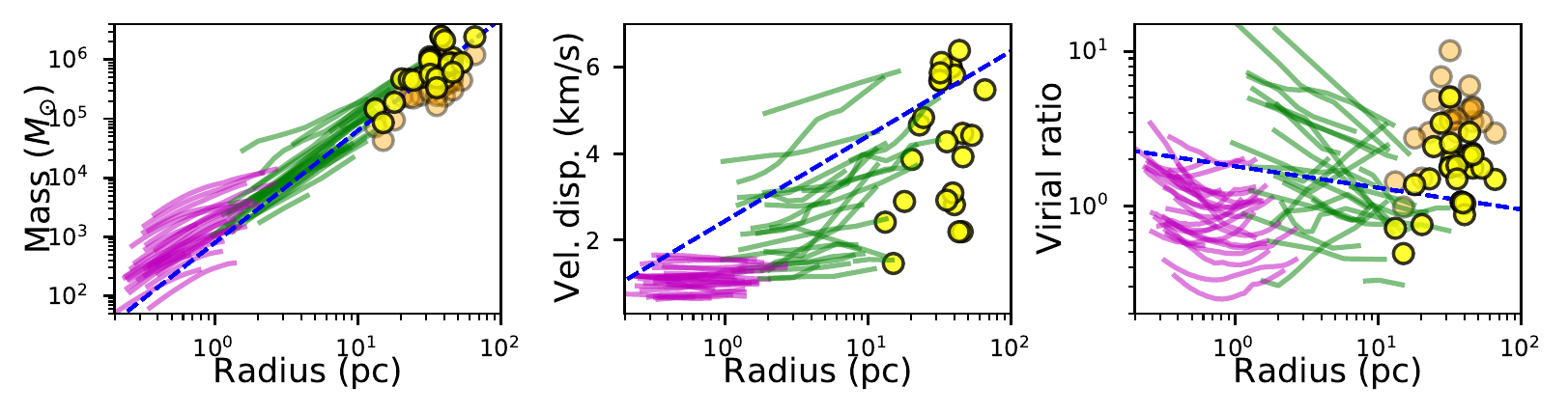} 
\vspace{-0.8cm}
      \caption{ The profiles are the same as those presented in the upper row of Fig.~\ref{obsprof}. In addition, we have added the $^{12}$CO(1-0) data points as presented in MD17 (transparent orange symbols) and once corrected by a factor of 2 in mass (yellow symbols). The blue dashed-lines show Larson's laws.}
         \label{prof_MD17}
   \end{figure*}

\begin{figure*}
  \vspace{-0.cm}
 \hspace{-.cm}
   \centering
   \includegraphics[width=18.cm]{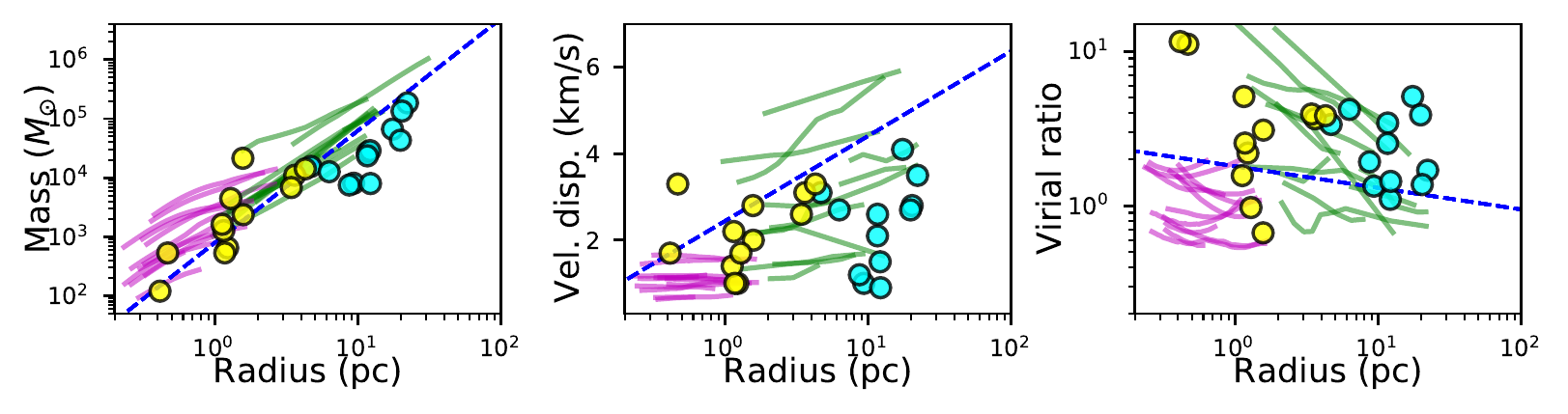} 
\vspace{-0.8cm}
      \caption{The profiles are the same as those presented in the upper row of Fig.~\ref{obsprof}, only restricted to the 12 clouds in common with H09. In addition, we have added the H09's measurements for these 12 clouds, for both radii measurements. We also overplot Larson's relations as dashed blue lines, and Solomon's size-velocity dispersion relation in the middle panel. }
         \label{sdcpropH09}
   \end{figure*}

How does this fit with studies such as that of \citet[MD17 hereafter]{miville-deschenes2017}, claiming that most molecular clouds are unbound? In order to answer that question we searched for the MD17 counterparts of all 24 molecular clouds  from our sample and compared the distributions of the virial ratio (as estimated by MD17) to the entire MD17 cloud population. Figure ~\ref{avirMD17} shows that our 24 IRDC-hosting molecular clouds are amongst the most gravitationally bound clouds from the MD17 sample, oversampling the low virial ratio tail of the distribution. So the fact that all the clouds studied in the present paper are gravitationally bound is not in contradiction with the MD17 results. We also notice that the virial ratio values plotted in Fig.~\ref{avirMD17}  and estimated by MD17 are larger than those we have estimated ourselves for the same sample of clouds. This difference could be real as MD17 computed their cloud properties from a lower gas density tracer, that is $^{12}$CO(1-0), or it could be due to systematics in the way properties are calculated. We investigated this by reporting the MD17 values of radius, mass, velocity dispersion, and virial ratio for all 24 clouds and added them to our observational profiles. Figure \ref{prof_MD17} (transparent orange circular symbols) shows an overall good agreement between out data points and those from MD17. However, while the radii reported by MD17 are larger, the masses are very similar to those we report on smaller radii. The much larger angular resolution of the data used by MD17 (i.e $8.5'$) means that their radii (and thus virial ratios) might be artificially increased compared to the values derived here by us. However, since $^{12}$CO(1-0) is a lower density tracer than $^{13}$CO(1-0), we do expect the MD17 counterparts to have larger radii. This therefore suggests that either we have overestimated our masses or MD17 have underestimated their $^{12}$CO masses. In MD17 they used a standard X$_{\rm{CO}}=2\times10^{20}$cm$^{-2}$ (K km$^{-1}$)$^{-1}$ factor to convert integrated $^{12}$CO intensities into H$_2$ column densities. As \citet{barnes2015} showed, this standard conversion factor typically underestimates column densities by a factor of $\sim2$ for resolved molecular clouds. Taking into account this change in X$_{\rm{CO}}$ would put the MD17 masses more in line with ours (see Fig~ \ref{prof_MD17} yellow circular symbols). In addition to this mass correction, one can wonder whether one should apply one to velocity dispersion measurements as well. Indeed, $^{12}$CO(1-0) is typically optically thick above H$_2$ column densities of few $10^{20}$~cm$^{-2}$, which means that mass and velocity dispersion measurements could be overestimated and underestimated, respectively. The effect on the velocity dispersion though is probably only of the order of 20\% \citep[e.g.][]{hacar2016}, but as the result of the $\sigma^2$ dependency of the virial ratio, a small correction factor on the velocity dispersion can lead to a significant difference on the virial ratios. However, as it can be seen in Fig.~\ref{prof_MD17}, the velocity dispersion measurements obtained by MD17 are in good agreement with ours, and we therefore do not believe that there is a systematic under-estimation of $^{12}$CO velocity dispersion for the clouds we are looking at. The corresponding distribution of virial ratios has also been reported onto Fig.~\ref{vir_comp} showing that 85\% of the $^{12}$CO-based virial ratio measurements are below 2.5, which is identical to the  $^{13}$CO-based virial ratio measurements. Overall, Figs.~\ref{prof_MD17} and ~\ref{vir_comp} show that even on scales of 100~pc, the vast majority of clouds from our sample have virial ratios that are consistent with being self-gravitating.

\begin{figure*}
  \vspace{-0.cm}
 \hspace{-.cm}
   \centering
   \includegraphics[width=18.cm]{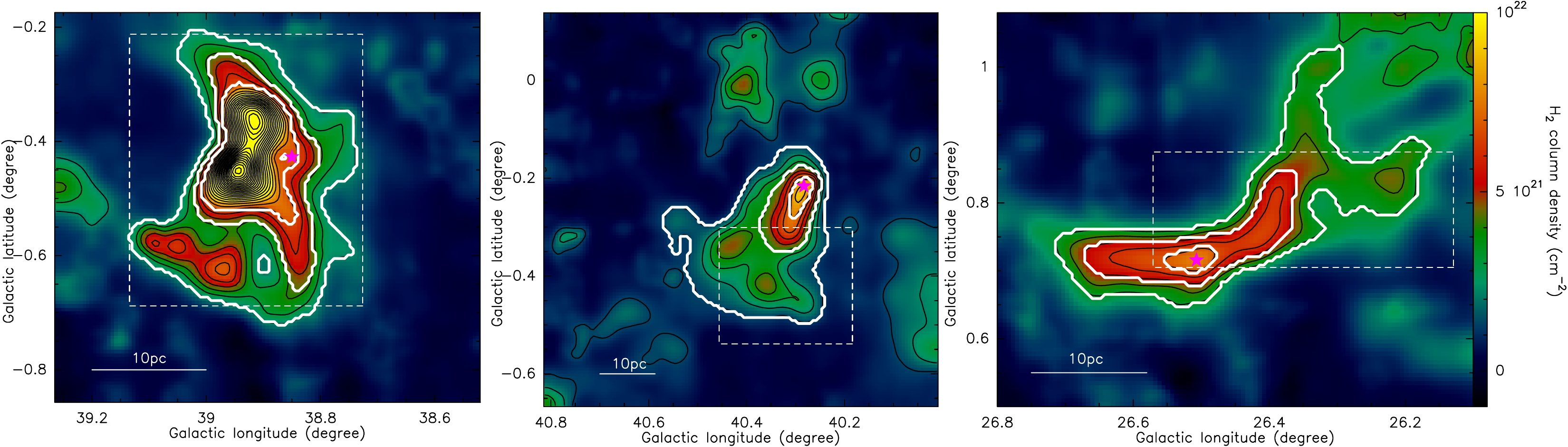} 
\vspace{-0.5cm}
      \caption{Colour and contours are the $^{13}$CO-based H$_2$ column density images of three of our clouds that are in common with H09's sample. The white dashed boxes show the area used by H09 to derive the large-scale masses represented as cyan symbols in Fig.~\ref{sdcpropH09}.}
         \label{H09boxes}
   \end{figure*}

\begin{figure*}
  \vspace{-0.cm}
 \hspace{-.cm}
   \centering
   \includegraphics[width=18.cm]{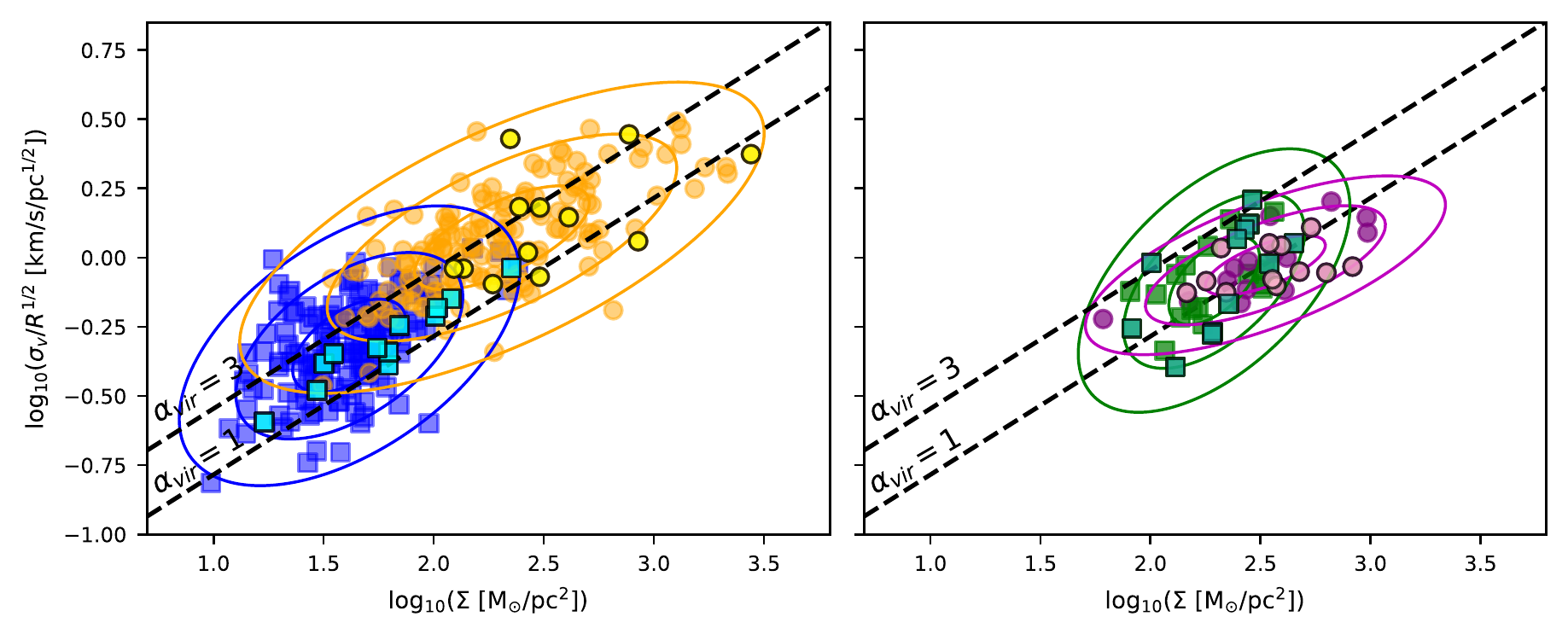} 
\vspace{-0.8cm}
      \caption{Heyer's plots (left): measurements from H09 on large (blue) and small (orange) scales. The points highlighted in cyan and yellow with black edges are those clouds in common with our sample; (right): Measurements from our study on large (green) and small (magenta) scales. The clouds in common with H09's sample are highlighted with light green and pink symbols with black edges. On each panel, the ellipses show the $1\sigma$, $2\sigma$, $3\sigma$  ellipses for each distributions, where $\sigma$ is the standard deviation. Lines of constant virial ratios of 1 and 3 are shown as black dashed lines.}
         \label{heyerplot}
   \end{figure*}

\subsection{Larson's, Solomon's, and Heyer's relations}

Probably one of the most influential studies on the observational characterisation of dynamical state of molecular clouds is that by Richard Larson in 1981. In that study, they found, mostly from using $^{13}$CO(1-0) data from the literature at the time, that the averaged cloud properties follow a number of relationships such that:   $\sigma\propto r^{\beta}$ with $\beta=0.38$, and $\rho\propto r^{-\gamma}$ with $\gamma=1.1$. A third relation, consequence of the first two, is that molecular clouds have virial ratios close to unity and $\alpha_{\rm{vir}}\propto r^{-\delta}$ with $\delta=0.14$. Larson's size-velocity dispersion relation has been interpreted as an evidence for turbulence-regulated gas dynamics, since $\beta=1/3$ is what one expects for incompressible Kolmogorov-like turbulence. However, these relations have later been revised by \citet[S87 hereafter]{solomon1987} who found a steeped size-velocity dispersion relation with  $\beta \simeq 0.5$. They suggested that such index is the direct consequence of the virialisation of individual molecular clouds at nearly constant mass surface densities. \citet[H09 hereafter]{heyer2009} reanalysed S87 cloud sample using $^{13}$CO(1-0) GRS data and determined that, even though cloud properties are compatible with being in virial equilibrium at all radii, the change in the internal mass surface density of clouds result in a different size-velocity dispersion relation to that proposed by Larson's and Solomon's. 

Finding out how our study compares to those mentioned above and understanding where the differences come from is fundamental if one wants to settle the question of the dynamical states of molecular clouds. Interestingly,  half of our cloud sample (12/24) is common to both H09 and S87's samples, and since H09 used the same $^{13}$CO data we use here, one can make a direct one-to-one comparison. The first property we compare is the distance used for all 12 clouds. As it can be seen in Fig.~\ref{distcomp}, for half of the clouds the distances match, while for the other half they do not. The latter group of clouds have been assigned the far distance by S87 and H09. Even though they have recalculated the kinematics distances, H09 have kept the near/far distance ambiguity solutions provided by S87. Looking in detail, these 6 clouds with far distances have been assigned so based on the fact that: i. they best fit the S87 size versus velocity dispersion relation; ii. They best match the scale-height of the molecular layer for that position and velocity range. These are both very questionable criteria. All clouds here host IRDCs, and it has been shown that 90\% of IRDCs are located at the near distance \citep{ellsworth-bowers2013}. This would suggest that maybe one of  the 12 clouds presented here is indeed at the far distance, but it is very unlikely that the 6 are. In the rest of the comparison we set the distance to all 12 clouds to that we give in Table 1.

\begin{figure*}
  \vspace{-0.cm}
 \hspace{-.cm}
   \centering
   \includegraphics[width=18.cm]{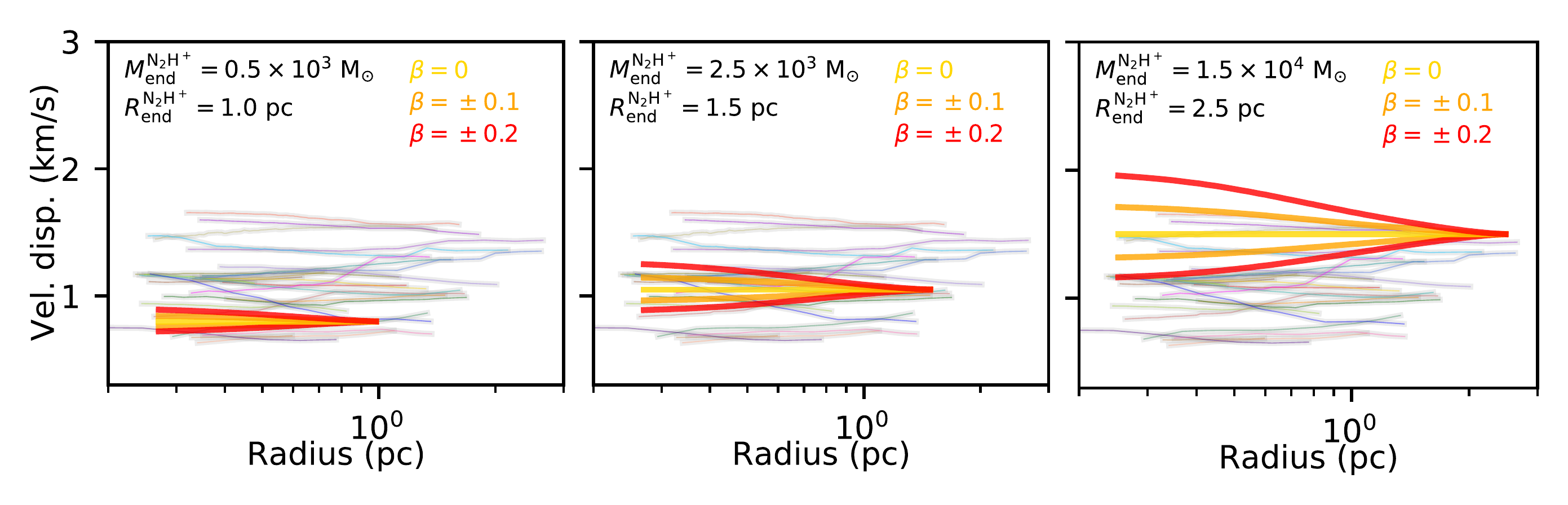} 
\vspace{-.8cm}
      \caption{Velocity dispersion profiles. The models displayed in each of the three panels correspond to three different clump masses and radii, but all with the same density profile $\gamma=2$. The yellow, orange, and red solid lines correspond to models with velocity dispersion profile index $\beta=0$, $\beta=\pm0.1$, and $\beta=\pm0.2$, respectively. The observed profiles are represented with thin coloured lines. }
         \label{betatest}
   \end{figure*}

 Figure~\ref{sdcpropH09} compares the profiles of the 12 clouds as we measured them with H09's measurements (after distance correction). In H09, each cloud has two measurements taken at different radii, both measured using $^{13}$CO. The cyan symbols represent the large scale measurements and the yellow symbols the small scale ones. Compared to our measurements of the same clouds, we can see that, at large radii, both H09's masses and velocity dispersions  tend to be underestimated. On small scales though, the masses are similar but the velocity dispersions are overunderestimateestimated. Before interpreting these discrepancies, one needs to understand the differences in the measurements themselves. For the large-scale mass measurements H09 used the original rectangular boxes that S87 used to measure their own masses. Those boxes where defined based the location and extent of the $^{12}$CO(1-0)  emission peaks derived from low-resolution high-noise maps. Figure~\ref{H09boxes} show three representative examples of such boxes overlaid on top of the cloud column density images. One can see that, with the exception of the biggest clouds, the boxes do not match the cloud morphologies, sometime missing the column density peaks, and often covering regions where no, or little, column density is present. The net impact of this is, for a given effective radius (defined as the radius of the disc having the same area as the box), the mass is heavily underestimated.  This problem mostly disappears for small scale mass measurements as H09 have for those used the contours of the column density maps (as we did). Regarding the velocity dispersion measurements on small scales, H09 overestimate them as a result of the same projection effect that is responsible for over-estimating our own $^{13}$CO velocity dispersion measurements. On large scale H09 underestimate the velocity dispersion most likely because of the unadapted velocity window used to compute their 1$^{\rm{st}}$ order moment. However, we cannot test this since velocity windows used for the integration by H09 are not provided. 

\begin{figure*}
  \vspace{-0.cm}
 \hspace{-.cm}
   \centering
   \includegraphics[width=18.cm]{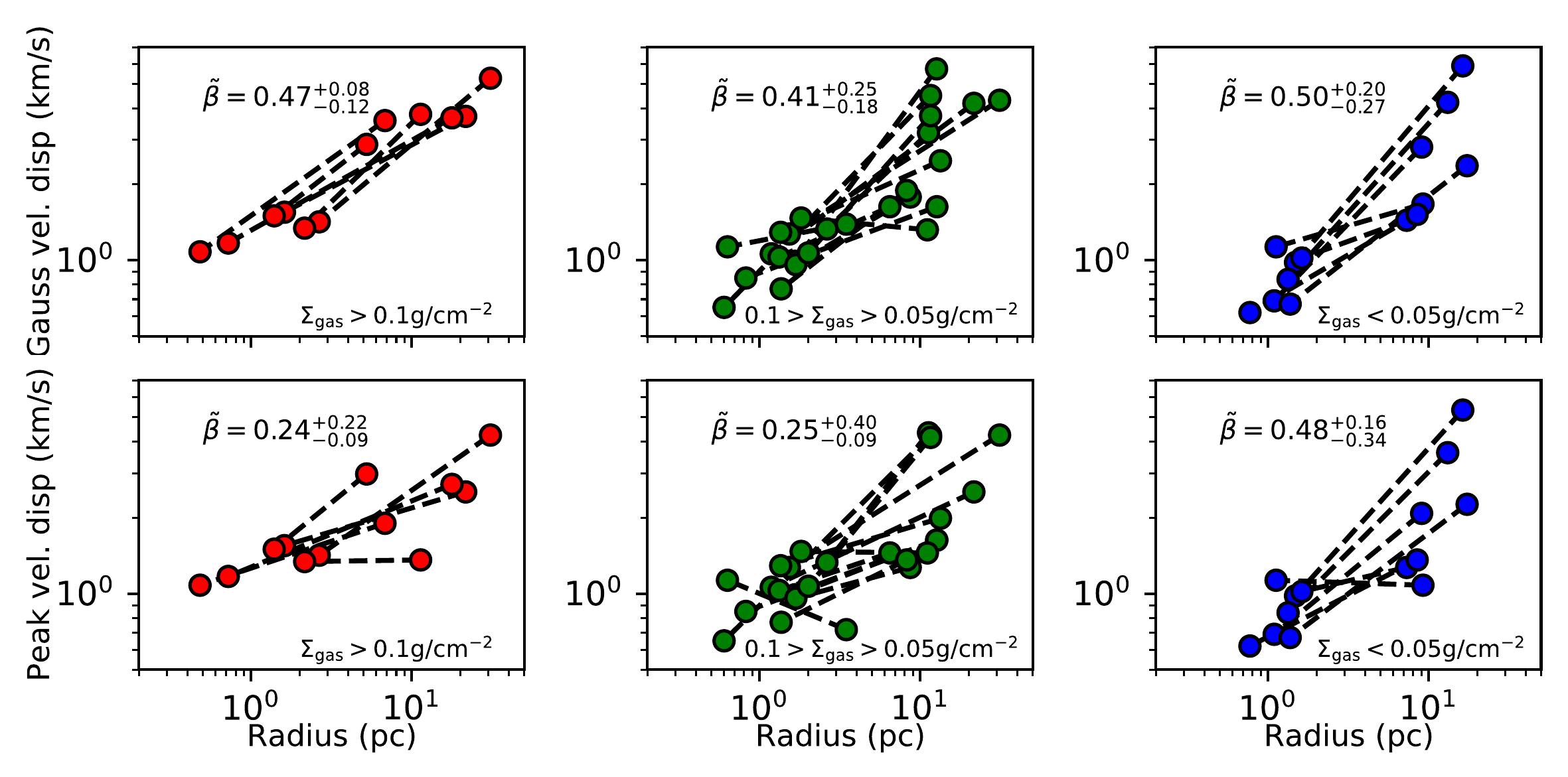} 
\vspace{-.8cm}
      \caption{Velocity dispersion profiles using only the measurements obtained on the largest clump and cloud scales (i.e. those quoted in Table 2). Clumps have been categories in three groups according to their average mass surface densities, each panel corresponding to one of those groups, with the range of corresponding mass surface densities being indicated. The top row corresponds to velocity dispersion measurements obtained from Gaussian fitting, the bottom row corresponds to velocity dispersion measurements obtained using the Peak method. Measurements from one clump-cloud pair are linked by a dashed black line. The median $\beta$ value, i.e $\tilde{\beta}$, and the corresponding  16$^{\rm{th}}$, and $84^{\rm{th}}$ percentiles are indicated in each panel.}
         \label{beta_masssurf}
   \end{figure*}

One particular plot that has been used by H09, and many others since, to support the picture of virialised clouds on all scales is one that plots the mass surface density $\Sigma_{\rm{gas}}$ of the clouds versus the parameter $p=\sigma_v/\sqrt{R}$.   On the left panel of Fig.~\ref{heyerplot} we reproduced the figure from H09, the clouds in common with our studies being highlighted with different colours (cyan and yellow) and with black edges. In this panel, it is quite clear that the large scale data points (the blue squares) are at lower mass surface densities than the small scales points (orange). The distribution of these points  stretch across more than two orders of magnitude along lines of constant virial ratios between 1 and 3.  On the same figure, the right-hand-side panel shows the same quantities for the common sample of clouds with properties as derived in this paper (here we used the values displayed in Table 2). We can see that the large scale points (green) completely overlap with the small scale points (magenta). When compared with Heyer's quantities for the same clouds we see that the spread is reduced by one order of magnitude.  This is a direct consequence of the measurement biases explained above. In fact, whether we look at the large scale measurements obtained on scales of 20~pc to 60~pc or measurements obtained on scales between 1.5pc and 5pc, the data points are located within a very similar area of the plot. This is a direct consequence of the density profile of the clouds being close to $\rho\propto r^{-1}$ on those scales and velocity dispersion profiles close to $\sigma\propto r^{0.5}$. 
 We also notice the quasi-absence of points below a 100\,M$_{\odot}$/pc$^{-2}$. As noted by \citet{schruba2019}, molecular clouds with lower mass surface density are non-self-gravitating. As our comparison with the MD17 virial ratio distribution shows, we are here biased towards the most self-gravitating clouds of the Milky Way population, it is therefore consistent to have nearly no measurements with mass surface density below 100\,M$_{\odot}$/pc$^{-2}$.

\subsection{Clump mass surface density versus $\beta$}

In their study of a sample of  29 clumps, \citet{traficante2020} found that the velocity dispersion profile index $\beta$ depends on the clump mass surface density. In that study, they determined that clumps at higher mass surface density (and mass) tend to have shallower velocity dispersion profiles (i.e. lower $\beta$ values) than low mass surface density clumps. The conclusion from that study was that the kinematics of  high mass surface density clumps is dominated by gravitational collapse leading to a departure from Larson's relation, believed to be driven by turbulence.

Figure~\ref{beta_masssurf} (top row) shows the velocity dispersion measurements obtained on the largest clump and cloud scales (Table 2) for each clump-cloud pair. Those measurements are obtained using the multiple Gaussian fit method. Clumps have been categorised into three groups according to their mass surface density following similar ranges as in \citet{traficante2020}. The median $\beta$ values given in each panel indicates that there is no significant difference between the three groups. However, the way \citet{traficante2020} have measured their velocity dispersion is different to what has been done here when using the Gaussian fit method. The main difference resides in the fact that they have performed a pixel-by-pixel analysis of  the $^{13}$CO(1-0) cube, by first clipping low signal-to-noise voxels, and then performing a 2$^{\rm{nd}}$ moment integration. As a result, the low intensity wings that we do detect, because of the spatial averaging, and fit are not represented in their velocity dispersion measurements. The closest velocity dispersion measurements we have made to those quoted in \citet{traficante2020} are those obtained with our Peak method (see Appendix B). As it can be seen in Fig.~\ref{beta_masssurf} (bottom row), using those Peak velocity dispersion measurement does change the picture.  Now we do see, albeit small number statistics, the same trend as that observed by \citet{traficante2020}, that is the highest mass surface density clumps have shallower velocity dispersion profiles. The main reason for the change in the median $\beta$ value is the decrease of the cloud-scale velocity dispersion measurements when using the Peak method (see Appendix B).

This leaves us with two possible interpretations regarding the differences observed between the two rows of Fig.~\ref{beta_masssurf}. The first one is that the high velocity wings that we fit using our multiple Gaussian fitting are unrelated to the clouds and that this method tends to overestimate the $^{13}$CO(1-0) velocity dispersion measurements. Note, however, that this overestimation would mostly be towards high mass surface density clumps since the low mass ones (right panels) remain mostly unchanged between the two methods. But if the high velocity wings were to be unrelated to the cloud of interest then there is no reason why one should observe a correlation between their presence and the clump mass surface density. The second interpretation is that those high velocity wings are truly associated to the clouds, and therefore the trend observed by \citet{traficante2020} between $\beta$ and the clump mass surface density is an artefact originating from their non-detection. While, in our view, this is the most likely interpretation, the comparison between the two methods still shows an interesting result in that high mass surface density clumps preferentially have parent clouds with highly complex kinematics. Whether complex velocity fields within molecular clouds are required for the formation of high mass surface density clumps, or whether high mass surface density clumps and their associated stellar feedback are responsible for generating complex velocity fields on larger scales remains to be understood.

\subsection{Dynamically decoupled clumps}

As the comparison of our observed profiles and spherical models have shown, the discontinuity in the observed velocity dispersion profiles is most likely the result of the combination of projection effects and a genuine change of the velocity dispersion profile index from $\beta\simeq0.5$ on large scales to $\beta\simeq0$ on small scales.  However, one can wonder how sensitive the observed velocity dispersion profiles are to the exact value of $\beta$ as the clumps only have a limited number of angular resolution elements in them. To test this, we built spherical models of varying $\beta$ index in order to set some constraints on the range of values compatible with our observations. Figure~\ref{betatest} displays the observed clump-scale velocity profiles along with different spherical models. Each panel corresponds to models of the same mass and radius, but with different velocity dispersion profiles. One can see that the different profiles are better resolved for the largest clouds, as expected. With this models in hands, it is also clear that $|\beta| < 0.2$ in all clumps, confirming the fact that the clump velocity dispersion profiles are flat and significantly different from Larson's profile.

\begin{figure}
  \vspace{-0.cm}
 \hspace{-.cm}
   \centering
   \includegraphics[width=8.5cm]{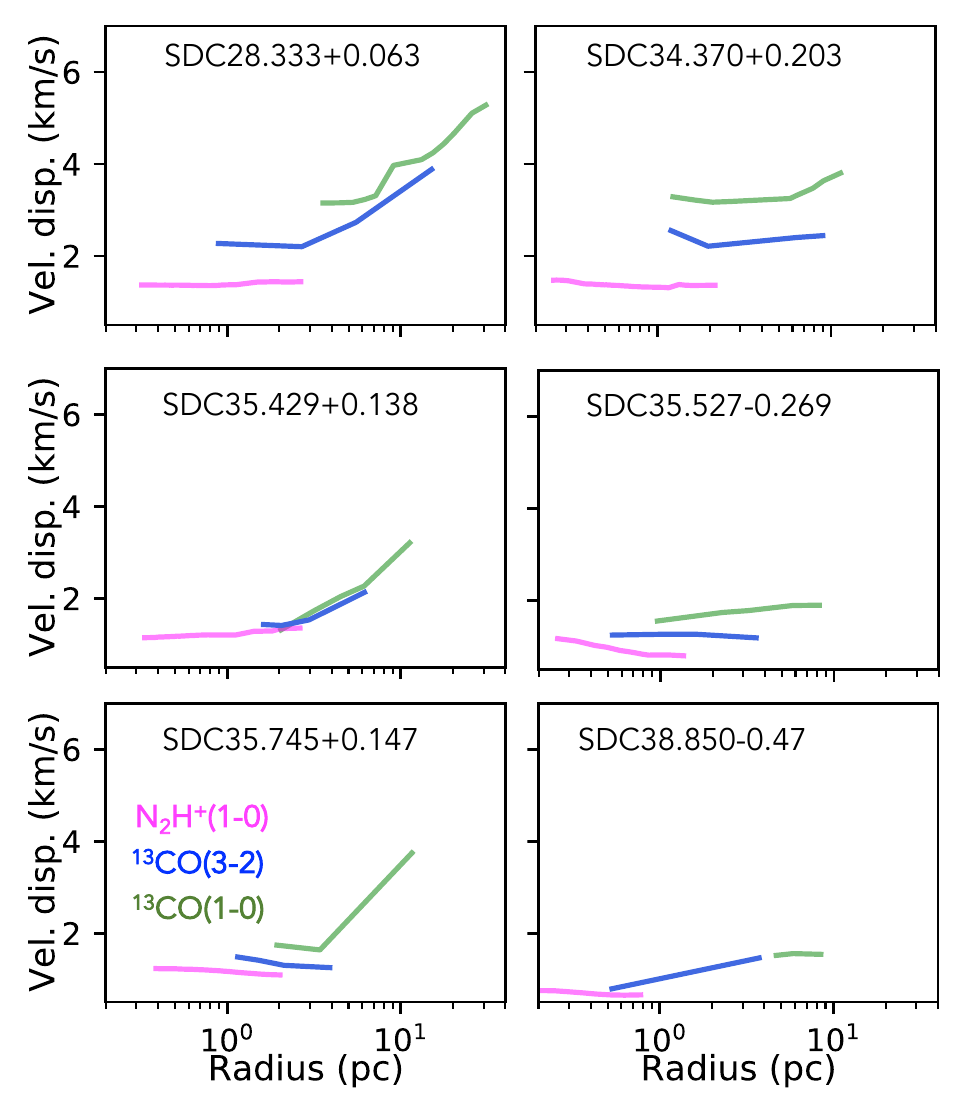} 
\vspace{-0.5cm}
      \caption{ Velocity dispersion profiles of a sub-sample of 6 clouds. The magenta and green lines are the same of those plotted in Fig.~\ref{obsprof} for those 6 clouds. The blue lines show the velocity dispersion profiles obtained from CHIMPS $^{13}$CO(3-2) emission. }
         \label{chimps}
   \end{figure}

The velocity dispersion discontinuity observed in Fig.~\ref{obsprof} between the N$_2$H$^+$(1-0) and the $^{13}$CO(1-0) measurements is, according to our models, the result of foreground/background layers of low-density and high-velocity dispersion gas that contaminate the $^{13}$CO(1-0) velocity dispersion measurements at small radii. If this interpretation of the observed profiles is correct, measuring the gas velocity dispersion with a line emission that traces intermediate gas densities should bridge, to some extent, the observed velocity dispersion discontinuity.
To test this conjuncture, we used the CHIMPS $^{13}$CO(3-2) survey data \citep{rigby2016}. Indeed, being a higher transition line, $^{13}$CO(3-2) is optically thinner and less extended than $^{13}$CO(1-0), making it a good tracer of intermediate gas densities. However, only 8 of our clouds have been covered by CHIMPS, amongst which one shows clear sign of self-absorption and has therefore been discarded. Figure~\ref{chimps} shows the velocity dispersion profiles of 6 of the 7 remaining clouds (one has been left out for a matter of figure readability). The $^{13}$CO(3-2) line has been fitted following the exact same procedure as for the $^{13}$CO(1-0) line. On this figure, we can see that the $^{13}$CO(3-2) velocity dispersion systematically lies at intermediate values between that of the other two tracers. Also, in most cases, the  $^{13}$CO(3-2) profiles nicely make the bridge between the denser and more diffuse gas. Altogether, these profiles further support our interpretation that clumps are dynamically decoupled from their parent molecular clouds. A sudden change in the velocity dispersion profile of the gas has been previously observed on core-scale, i.e. $\sim0.1$pc \citep[e.g.][]{pineda2010}. Whether or not this core-scale transition to coherence has the same physical origin to the proposed clump dynamical decoupling remains to be shown.

As discussed in Sec.~5.1 the vas majority of the clouds studied here are consistent with being self-gravitating on all scales. This means that the observed change in the velocity dispersion profiles is unlikely to be the result of the gas switching from a non self-gravitating state to a self-gravitating state. When a cloud is self-gravitating, it can only be in two states: either it is collapsing, or in quasi-static equilibrium. One possibility would be for instance that the clouds from our sample, as proposed by \citet{vazquez-semadeni2019}, are collapsing on all scales. However, the collapse in these models is scale-free, with no transition regime at any scale. It is important to note though that  protostellar outflows \citep[e.g.][]{cabral2012,hsieh2023} are not included in such simulations, and could play a role in generating a change in the velocity dispersion profiles of collapsing clouds. It could also be that, as the specific angular momentum increases during the collapse, clumps become somewhat supported by rotation \citep[e.g.][]{lee2016a}. However, this seems incompatible with a steeper clump density profile, and no systematic observation of rotation motion is observed in these clumps (Peretto et al. in prep.). A third and preferred possibility is that clouds are stable on the largest scales and that they collapse on clump scale. Indeed, both density ($\gamma=2$) and velocity dispersion profiles ($\beta=0$) derived from our 1D modelling of the clumps are asymptotic solutions to a spherical isothermal non-free-falling collapsing cloud with initial uniform density \citep{larson1969,penston1969}, and as noted by these authors the self-similar nature of the solution means that it may apply to any structure (i.e. protostellar core, clump, cloud). Note however that what we observe is the velocity dispersion profile and not the infall velocity profile. Even though we do expect a relationship between the two, it is not clear whether both are expected to have the exact same index. Through recent analytical models, although with different settings,  both \citet{guang-xi2018} and  \citet{gomez2021} show that $\gamma=2$ naturally arises from the gravitational collapse of cores and clumps. There is also now plenty of evidence for clump collapse and clump accretion \citep[e.g.][]{peretto2006,peretto2007,peretto2013,peretto2014,peretto2020,schneider2010,traficante2018,williams2018,schworer2019, barnes2019,anderson2021,rigby2021,bonne2022,zhou2022,xu2023}. A possible agent that might be able to stabilise the clouds on the largest scales is stellar feedback. For instance, in \citet{watkins2018}, it has been shown that stellar feedback from embedded O stars does not impact much the dynamical properties of the dense gas that has already been assembled, but does clearly modify the structure of the larger scale clouds. This is compatible  with the observed change in velocity dispersion profiles presented here. Even though most clumps in our study do not have any embedded H$_{\rm{II}}$ regions associated to them, injection of momentum and energy within the more diffuse cloud could come from nearby sites of massive star formation. A picture in which gas collapses on clump scale while being supported by turbulence on larger scales is consistent with the model proposed by \citet{guang-xi2017}.

\begin{figure}
  \vspace{-0.cm}
 \hspace{-.cm}
   \centering
   \includegraphics[width=8.5cm]{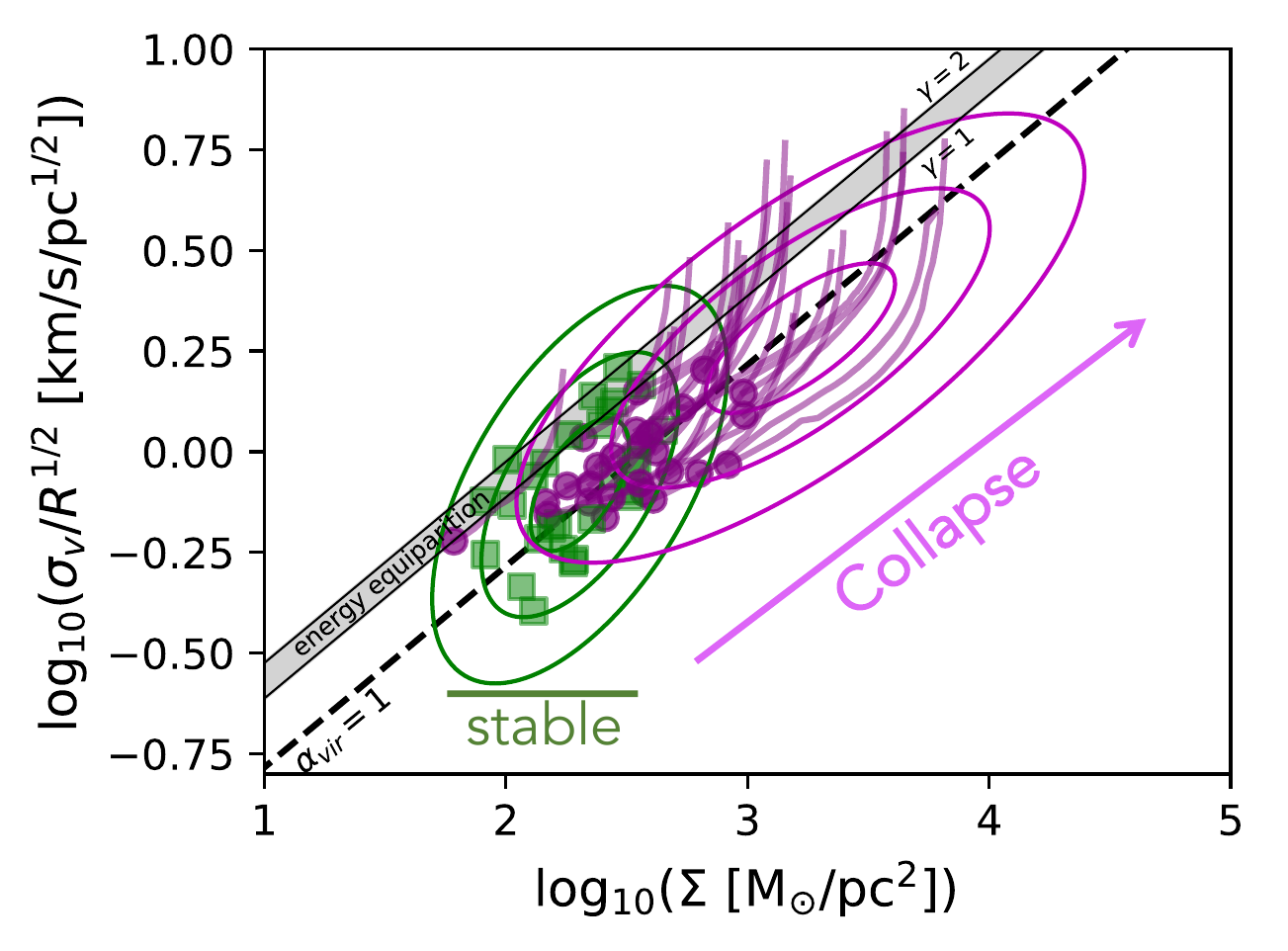} 
\vspace{-0.7cm}
      \caption{Same as Fig.~\ref{heyerplot} with the addition of all measurements made from N$_2$H$^+$(1-0) (purple lines). The shaded area show the region of energy equipartition for density profile indices between $\gamma=1$ and $\gamma=2$. }
         \label{heyerplot_collapse}
   \end{figure}

 Another possible agent that could stabilise the cloud is magnetic field. An increasing number of studies suggest that magnetic fields are dynamically important/dominant in the low density regions of molecular clouds. A transition in the relative orientation between magnetic field and the density gradients of interstellar structures has been interpreted as evidence for a change in the dominant energy source, from magnetic energy on large scale to gravitational energy on clump scale \citep[e.g.][]{soler2013,soler2016,chen2016,planck2016a,planck2016b,tang2019,arzoumanian2021}. In Vela C, \citet{fissel2019} determined that this change of relative orientations occur at a number density of $n\sim10^3$~cm$^{-3}$. The change in the velocity dispersion profiles we observe in our sample could then be the dynamical counterpart of that "magnetic" transition.  Figure~\ref{heyerplot_collapse} shows the same plot as in Fig.~\ref{heyerplot} in which the largest cloud scale measurements and all the clump scale measurements are shown. On that plot it becomes obvious that clumps behave differently than their parent clouds. The clump mass surface densities increase over two order of magnitudes along lines of constant virial ratios except towards the most central points where $p$  and the virial ratios increase. To our knowledge no theoretical equivalent to the plots we are producing here exists. However, a scenario in which parsec-scale clumps are collapsing while their parent molecular clouds are in quasi-static equilibrium seems to intuitively match what we see.

An important aspect of Fig.~\ref{obsprof}  is the relatively large range of velocity dispersion profile indices on cloud scale (the green lines). They range from being flat $\beta\sim0$ to relatively steep $\beta>0.5$. The fact that molecular clouds of several $10^4$~M$_{\odot}$ on tens of parsec-scale may have velocity dispersions that are barely above 1~km/s clearly shows that Larson's relation is just a statistical average over clouds of very different dynamical states.  In fact, it is possible that the range of velocity dispersion profiles correspond to different evolutionary stages in the formation and evolution of molecular clouds and clumps within. Any scenario that attempt to explain the dynamical decoupling of clumps needs to do so in the context of this observed variety of large-scale velocity dispersion profiles.

\section{Summary and conclusion}

We performed the analysis of 27 IRDC embedded within 24 molecular clouds. We computed the mass, velocity dispersion, and virial ratio profiles of each of them using three different datasets: {\it Herschel}-derived H$_2$ column density maps, GRS $^{13}$CO(1-0)-derived H$_2$ column density cubes, and N$_2$H$^+$(1-0) data cubes. The combination of these data allowed us to probe both the dense and diffuse parts of the clouds, with radii from $\sim0.2$\,pc up to $\sim30$\,pc. Using 1D power-law models we can explain the origin of the different features observed in those profiles and we conclude that: 1. the vast majority of cluster-forming molecular clouds are consistent with being self-gravitating on all scales; 2. the diffuse part of the cloud has a shallow density profile ($\gamma\sim1$) that steepens ($\gamma\sim2$)  in the densest parts on a couple of parsec scale; 3. the velocity dispersion profile switches, for most clouds, from $\beta\sim0.5$ in the diffuse part of the clouds to $\beta\sim0$ in the denser parts. We discuss the possible interpretation of such a decoupling of the clumps from their surrounding cloud and conclude that the observations are best explained by a universal global collapse of dense clumps embedded within stable molecular clouds, even though we cannot completely rule out a scenario in which the entire cloud collapses, with small-scale feedback, such as protostellar outflows, impacting the gas kinematics on clump scales.  We also notice that the velocity dispersion profiles on molecular cloud scales (i.e. $>2$~pc) show a large variety of $\beta$ values, some very far from the standard Larson's relation, which might be linked to their evolution since the time of their formation.

Understanding the origin of the observed low star-formation efficiency (SFE) in molecular clouds is one of the main goals of star formation research.  A low SFE involves a scale/density-dependent dynamical state of the gas in which most of a cloud mass is not directly involved in the formation of stars. Observationally, the existence of a star formation threshold has been discussed in the context  of the study of nearby star-forming clouds \citep[e.g.][]{lada2010,heiderman2010, pokhrel2020}. However, so far, no studies has searched for direct evidence of a transition regime in the dynamical properties of the gas within individual molecular clouds. The work presented here clearly suggests that such transition regime does exist. Because parsec-scale clumps are believed to be the direct progenitors of star clusters \citep[e.g.][]{krumholz2019}, our results hence suggest that star cluster formation is not a scale-free process.

Our results also carry a number of key questions and implications. First, we here do not explain what the trigger of the clump collapse is, whether it is the result of a gravitationally instability or the diffusion of magnetic fields, or any other mechanism. We also do not explain what is the main agent that counter-balance gravity in the diffuse parts of the clouds. These questions will have to be answered if one wants to derive a comprehensive scenario for the formation of star clusters. Also, one implication of our results is the fact that star formation is likely to be mostly confined to these parsec-scale collapsing clumps. Therefore their properties define the initial conditions for cluster formation, and understanding the link, on one side, between the properties of clumps and that of their associated protostellar population, and on the other side, between the global population of Galactic clumps and the star formation rate and efficiency of the Milky Way remains a fundamental challenge. 

\section*{Acknowledgements}

 We would like to thank the referee, Erik Rosolowsky, for a careful and well balanced report that helped improving the quality of the paper.
 NP and AJR acknowledges the support of STFC consolidated grant number ST/N000706/1 and ST/S00033X/1.
FL acknowledges support by the Marie Curie Action of the European Union (project \textit{MagiKStar}, Grant agreement number 841276). G.A.F acknowledges support from the Collaborative Research Centre 956, funded by the Deutsche Forschungsgemeinschaft (DFG) project ID 184018867. G.A.F also acknowledges support from the University of Cologne and its Global Faculty Program.  This work is based on observations carried out under project number 023-13 with the IRAM 30m telescope. IRAM is supported by INSU/CNRS (France), MPG (Germany) and IGN (Spain).

\section*{Data Availability}

 The {\it Herschel} and GRS data used in this article are already publicly available on their respective survey webpages. The IRAM 30m N$_2$H$^+$(1-0) data can be provided upon reasonable request to N. Peretto.



\bibliographystyle{mnras}
\bibliography{references} 




\appendix

\section{Summary figures}

This appendix is supplied as online supplementary material.

\section{Velocity dispersion: method comparison}

\begin{figure*}
  \vspace{-0.cm}
 \hspace{-.cm}
   \centering
   \includegraphics[width=18.cm]{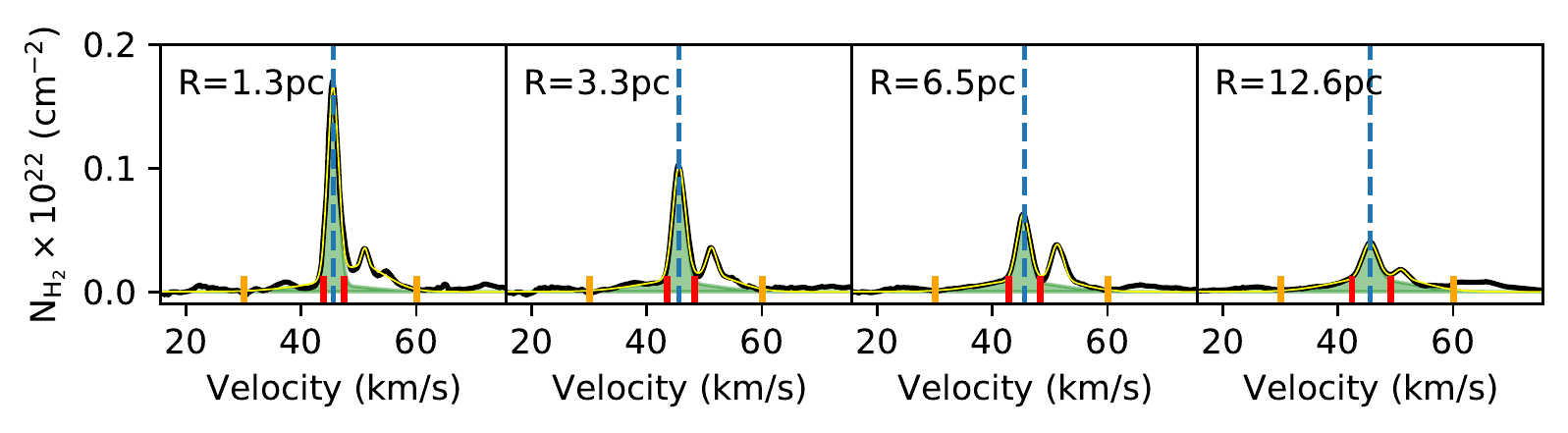} 
\vspace{-0.0cm}
      \caption{Examples of $^{13}$CO-based H$_2$ column density spectra for SDC18.624-0.070 in black. The multi Gaussian fits are shown as thin yellow lines, the shaded green area correspond to the components we believe are associated to the cloud. The velocity intervals for the moments method are shown as vertical orange ticks, while the velocity intervals for the Peak method are shown as vertical red ticks. }
         \label{velrange}
   \end{figure*}

\begin{figure*}
  \vspace{-0.cm}
 \hspace{-.cm}
   \centering
   \includegraphics[width=18.cm]{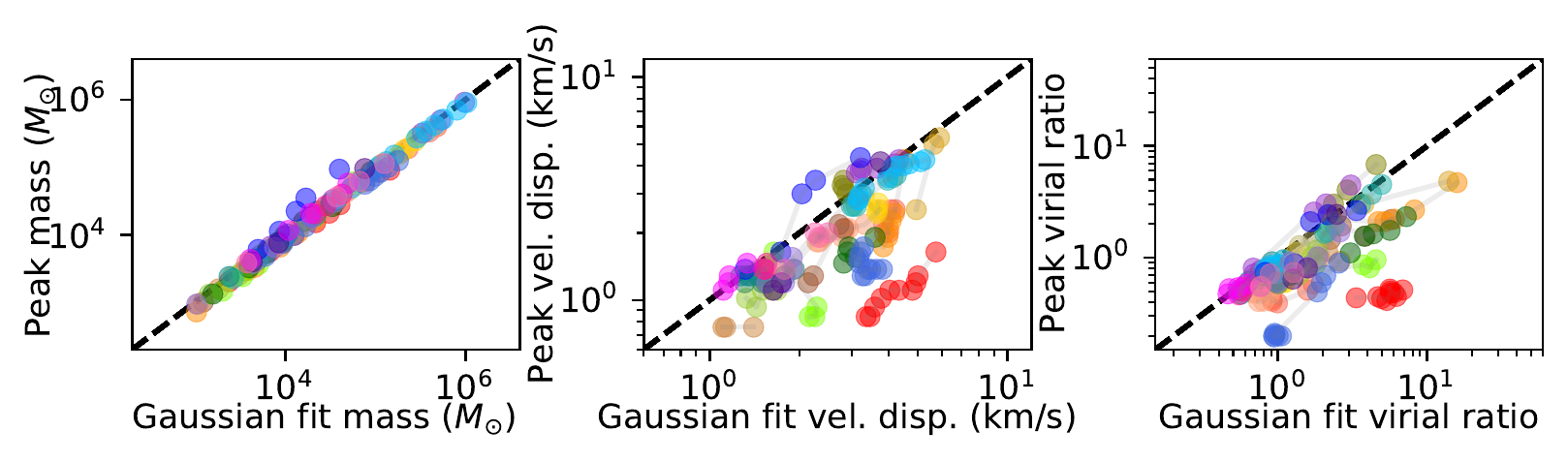} 
    \includegraphics[width=18.cm]{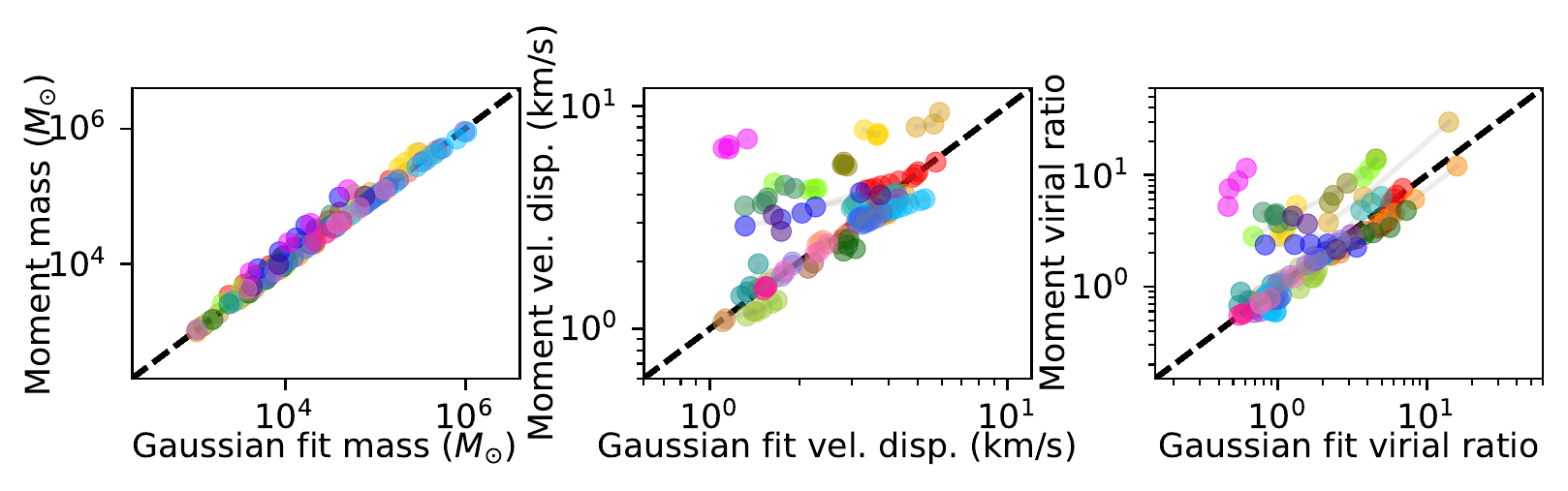} 
\vspace{-0.0cm}
      \caption{Comparing cloud properties using different methods for estimating the velocity dispersion of $^{13}$CO(1-0). In the top row we compare the Gaussian fit results on the x-axis, the method used in the rest of the paper, against the Peak results on the y-axis. In the bottom row, the x-axes remain the same, but the y-axis show the Moments results. The black dashed lines represent the one-to-one relationships, and each colour corresponds to a different cloud, each cloud having a number of measurements taken at different radii.}
         \label{compmeth}
   \end{figure*}

In addition to the Gaussian fits presented in the paper, we also tested two other methods that are often used in the literature. The first of those  is a traditional moments method, referred to as Moments in this Appendix, for which we compute the 0$^{\rm{th}}$, 1$^{\rm{st}}$, and 2$^{\rm{nd}}$ order moments within a given velocity interval. The 0$^{\rm{th}}$ order moment is used to compute the mass. The 1$^{\rm{st}}$ and 2$^{\rm{nd}}$ order moments are used to compute the velocity dispersion. The velocity intervals are determined by eye and are defined so that the cloud main component and any overlapping emission from overlapping clouds, in velocity space, are included (see Fig.~\ref{velrange} for an example). The second method we tested, referred to as Peak in this Appendix, is based on the full-width-at-half-maximum (FWHM) of the column density spectrum peak. The velocity dispersion is taken as FWHM/2.35, and the velocity interval over which the mass is calculated is taken as eight times the dispersion, and is centred in a way that matches the  asymmetry of the peak velocity with respect to the FWHM velocity interval  (see Fig.~\ref{velrange}). To obtain the mass we just integrate the spectra over that velocity range.

Figure~\ref{compmeth} compares the resulting masses, velocity dispersions, and virial ratios as obtained for the three methods (Gaussian fit, Moments, and Peak). The top row shows the comparison between the Gaussian fit and Peak methods, while the bottom row shows the comparison between the Gaussian fit and the Moments method. This shows that, as far as the masses are concerned, the method used does not make much of a difference, and the reason is that most of the mass is located within the central few channels that are covered by all three methods. The main differences between the methods are related to the estimate of the velocity dispersions. One can see that the Peak method nearly systematically produces lower velocity dispersions than the Gaussian fit method. This is expected as that method only focusses on the central peak of emission and will therefore exclude any large velocity dispersion wings that might be present. The results of the Peak method are similar to what we would get by fitting the spectra with single Gaussians. When comparing the Moments results to those of the Gaussian fits one can see that they are, overall, in broader agreement, even though the Moments method tend to produce larger velocity dispersions for some clouds. Those are the clouds for which a spectrally overlapping components has been excluded from our Gaussian fit results but included within the Moments one (as for SDC18.624-0.070 presented in Fig.~\ref{velrange}). The virial ratios differences are a direct consequences of the differences observed in the velocity dispersion measurements, leading to virial ratios than can be as high as $\sim50$ in some cases when using the Moments method and as low as $\sim0.1$ when using the Peak method.

\begin{figure*}
  \vspace{-0.cm}
 \hspace{-.cm}
   \centering
   \includegraphics[width=18.cm]{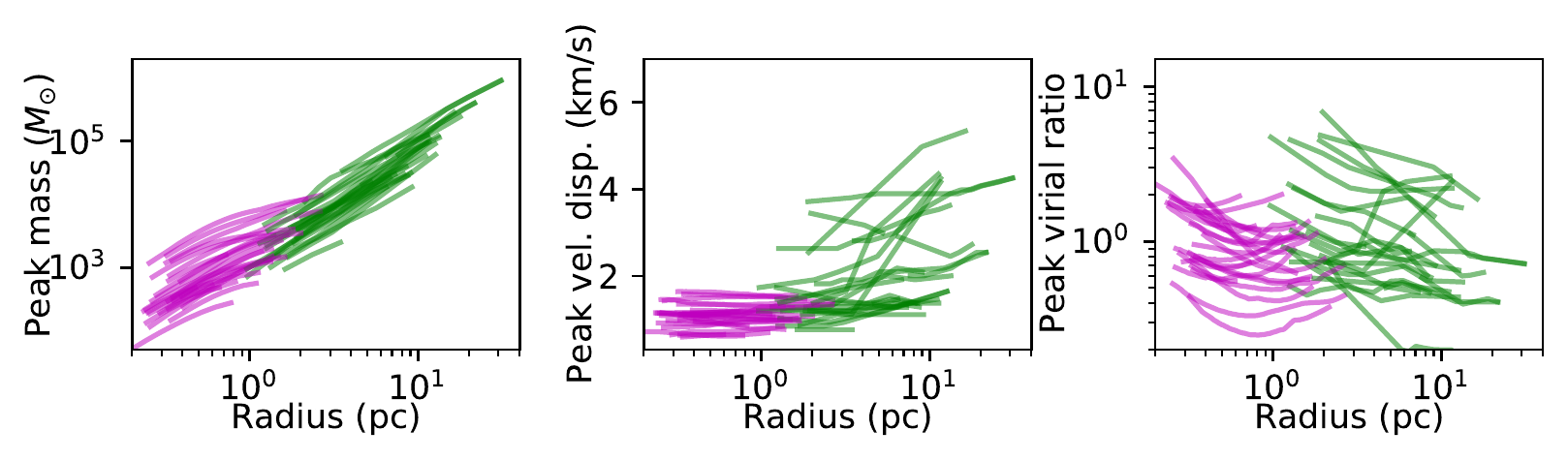} 
     \includegraphics[width=18.cm]{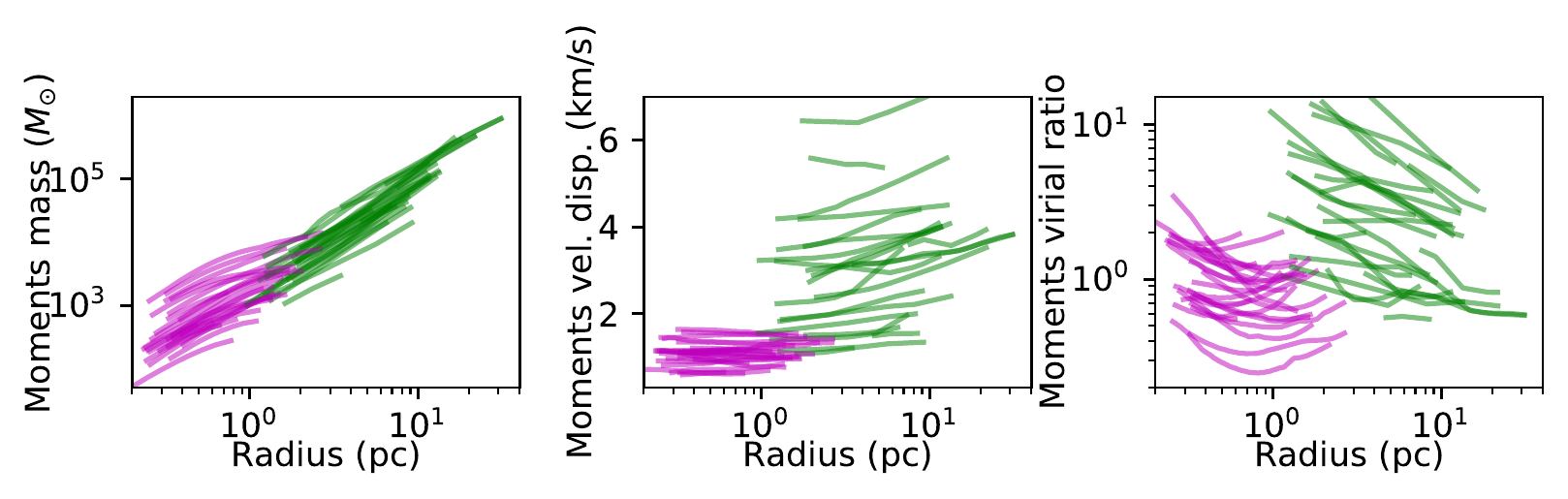} 
\vspace{-0.5cm}
      \caption{Same as Fig.~\ref{obsprof} but with the mass and velocity dispersion estimated via the Peak (top) and Moments (bottom) methods.}
         \label{PeakProf}
   \end{figure*}

 For completeness, we also plot the mass, velocity dispersion, and virial profiles as obtained for both the Moments and Peak methods (see Fig.~\ref{PeakProf}).  As expected from our previous discussion, the main differences lie in the velocity dispersion profiles, whereby the Moments method produce a larger discontinuity between clump and cloud scales, while the Peak method makes the discontinuity less prominent. However, as our models show (Figs.~\ref{model_prof} and \ref{model_prof_broken}) for shallow cloud density profiles as those observed (i.e. $\gamma\le1.5$) significant velocity discontinuities are expected, which indicates that the Peak method is likely underestimating the  true gas velocity dispersion.

Overall, we believe that the Gaussian fits method provide better results than any of the other two as it allows to include large velocity dispersion wings and exclude, at the same time, components that we know are not physically related to the cloud of interest. It is definitely the best method to measure the velocity dispersion of $^{13}$CO clouds.

\section{The $N_{\rm{N_2H^+}}^{\rm{edge}}=0$ case}

\begin{figure}
  \vspace{-0.cm}
 \hspace{-.cm}
   \centering
   \includegraphics[width=8.5cm]{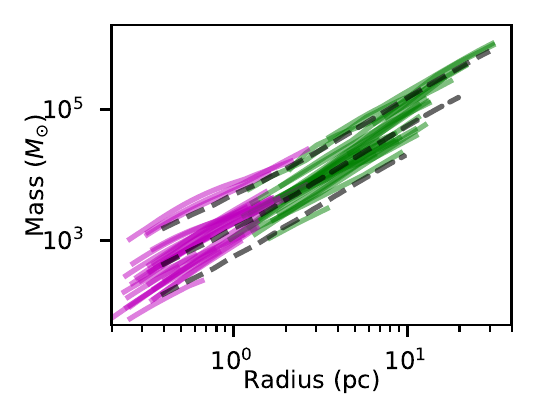} 
\vspace{-0.5cm}
      \caption{Same as the mass profiles displayed in Fig.~\ref{obsprof} but for the case where $N_{\rm{N_2H^+}}^{\rm{edge}}=0$ (i.e. no background subtraction for the clump-scale masses). The three black dashed lines show three $\gamma_{\rm{in}}=2$ and $\gamma_{\rm{out}}=1$ models of three different masses and external radii. These models are not convolved in order to maximise the visibility of the slope change.}
         \label{massprof_nobg}
   \end{figure}

In this Appendix, we quantify the impact of not removing a background column density on the clump-scale mass profile, effectively setting $N_{\rm{N_2H^+}}^{\rm{edge}}=0$. Figure ~\ref{massprof_nobg} shows the corresponding mass profiles. One can see that the profiles, as expected, appear a lot more continuous than when a background is removed. Interestingly, there is no obvious change of mass profile slope, which seems to be in apparent contradiction with our previous claim that clumps have a steeper density profile than their parent molecular cloud. In order to check whether or not we would be able to observe such a slope change, we produced a series of three models (See Sect.~4) with $\gamma_{\rm{in}}=2$ and $\gamma_{\rm{out}}=1$ of different masses. Those three models are represented as black dashed lines in Fig.~\ref{massprof_nobg}. These models clearly show that as a result of line-of-sight mass contamination one is unable to observe any significant change in the mass profile slope, i.e. the measured mass at clump scale is dominated by the cloud-scale foreground/background mass (see also Fig.~\ref{prof_ratios}). Therefore, the apparent contradiction is not one, and background subtraction is necessary if one wants to evidence any change in the radial density profile power-law index.

\section{Models of bijective mass and velocity dispersion estimates}

Here we describe how we built the projected version of the spherical models presented in the paper. 

First, let us consider a sphere of radius with the following the following density profile:

\begin{equation}
\rho(r)=\rho_0\left(\frac{r}{r_0}\right)^{-\gamma}
\end{equation}

\noindent where $r_0$ and $\rho_0$ are normalisation constants. Then the enclosed mass within radius $r$, $m_{\rm{real}}(r)$, is given by:

\begin{equation}
m_{\rm{real}}(r)=\int_0^r\rho_0\left(\frac{r}{r_0}\right)^{-\gamma}4 \pi r^2dr
\end{equation}

\begin{equation}
m_{\rm{real}}(r)=4\pi\rho_0 r_0^{\gamma} \frac{r^{3-\gamma}}{3-\gamma}     
\end{equation}

\begin{figure}
  \vspace{-0.cm}
 \hspace{-.cm}
   \centering
   \includegraphics[width=6.cm]{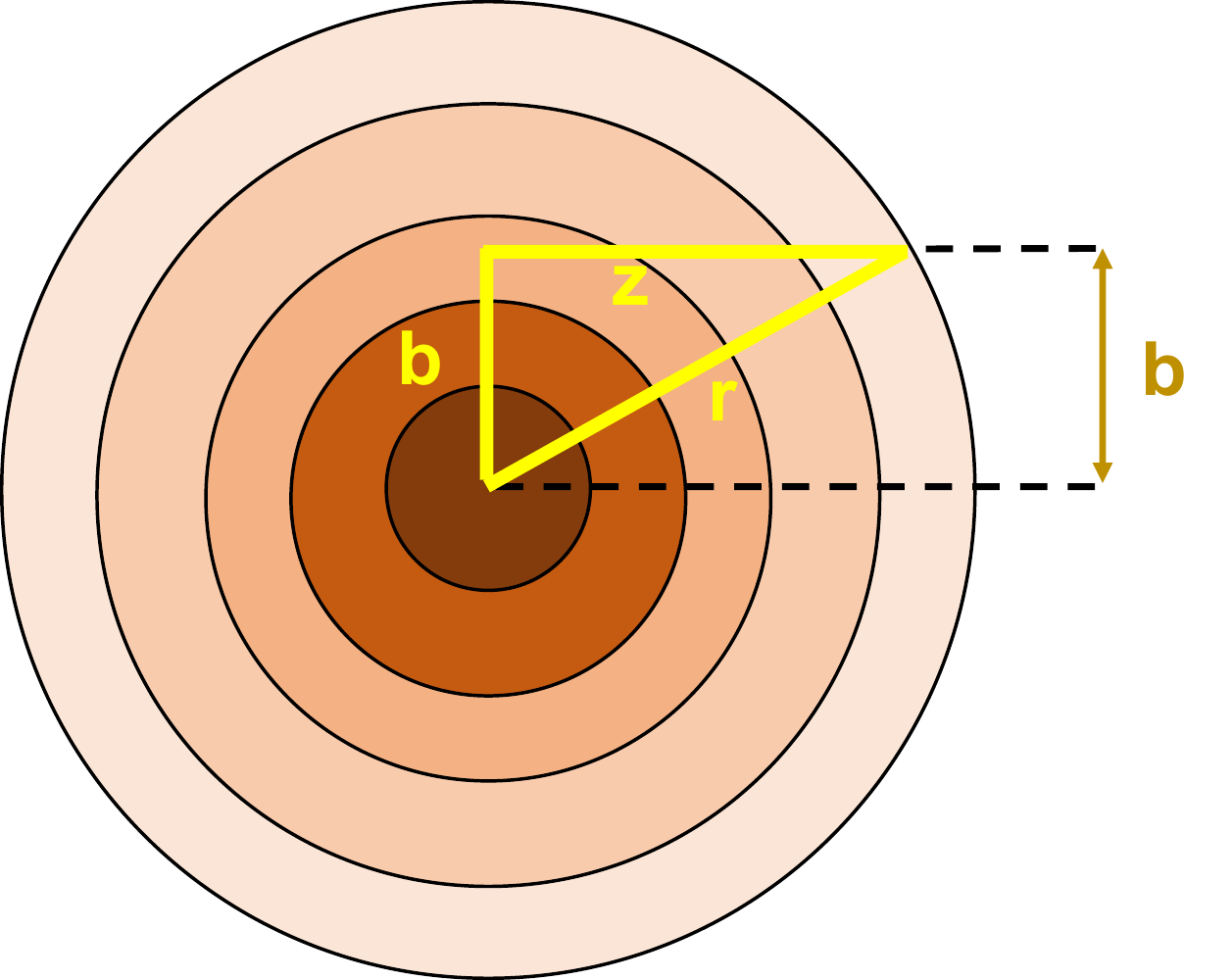} 
\vspace{-0.0cm}
      \caption{Sketch illustrating the definition of the different variables.}
         \label{sketch3}
   \end{figure}

\noindent Now we need to derive the observed mass, $m_{\rm{obs}}(r)$, derived from the observed column density map using the bijection method.  The column density is obtained by:

\begin{equation}
N(b)=2\int_0^{z_{\rm{max}}}\rho(z)dz
\label{col1}
\end{equation}

\noindent where $b$ is the impact parameter, equivalent to the projected radius, $z$ is the distance along the line of sight from the tangent point of the sphere of radius $b$, and $z_{\rm{max}}$ is the maximum distance along the line of sight from the tangent point to the edge of the cloud. Figure~\ref{sketch3} summarises these different variables.  The variable $z$ can be expressed as:
\begin{equation}
z=\sqrt{r^2-b^2}
\end{equation}

\noindent with its maximum value $z_{\rm{max}}$ being:

\begin{equation}
z_{\rm{max}}=\sqrt{R_{\rm{cloud}}^2-b^2}
\end{equation}

\noindent where $R_{\rm{cloud}}$ is the cloud radius. Equation~\ref{col1} can then be written as:

\begin{equation}
N(b)=2\rho_0r_0^{\gamma}\int_0^{z_{\rm{max}}}\left(b^2+z^2\right)^{-\gamma/2}dz
\label{col2}
\end{equation}

\noindent  The observed mass within a projected radius $b=r$ is given by:

\begin{equation}
m_{\rm{obs}}(r)=2\pi \int_0^{r} N(b) b db 
\label{mobs}
\end{equation}

\noindent  Equation~\ref{col2} can be solved analytically for some specific values of $\gamma$, but can easily be solved numerically for any value of $\gamma$ as long as $0\le \gamma <3$.

Now, let us consider that the same sphere has the following velocity dispersion profile:
\begin{equation}
\sigma(r)=\sigma_0\left(\frac{r}{r_0}\right)^{\beta}
\end{equation}

\noindent where $\sigma_0$ is a normalisation constant. The corresponding mass-weighted velocity dispersion is given by:  

\begin{equation}
\overline{\sigma}_{\rm{real}}(r)=\frac{\int_0^r\sigma(r)\rho(r)4\pi r^2 dr}{m_{\rm{real}}(r)}
\end{equation}

\begin{equation}
\overline{\sigma}_{\rm{real}}(r)=\frac{4\pi\sigma_0\rho_0r_0^{\gamma-\beta}}{m_{\rm{real}}(r)}\frac{r^{3-\gamma+\beta}}{3-\gamma+\beta} 
\end{equation}

\noindent Now, the observed velocity dispersion $\overline{\sigma}_{\rm{obs}}$ is given by:

\begin{equation}
\overline{\sigma}_{\rm{obs}}(b)=\frac{\int_0^{z_{\rm{max}}}\rho(z)\sigma(z)dz}{N(b)}
\label{sig1}
\end{equation}

\begin{equation}
\overline{\sigma}_{\rm{obs}}(b)=\rho_0\sigma_0r_0^{\gamma-\beta}\frac{\int_0^{z_{\rm{max}}}(b^2+z^2)^{(\beta-\gamma)/2}dz}{N(b)}
\label{sig2}
\end{equation}

\noindent Equation \ref{sig2} can easily be numerically integrated for different combinations of $\beta$ and $\gamma$ values. Note that here, we derived the expression of $\overline{\sigma}_{\rm{obs}}(r)$ and $m_{\rm{obs}}(r)$ in the case of single power law profiles, but, when extended to broken power laws, the expressions becomes longer as one integration has to be made
for each part of the profile. The logic behind it being the same as for single power law profiles, we will not derive their expressions here.

\section{Impact of non-uniform velocity dispersion profiles on the virial ratio parameter}

As noted by MD17 the non-uniformity of the velocity dispersion within clouds impact their virial ratio estimates. Here we show the derivation of the corresponding correcting factor, noted $a_3$ in MD17. 

The kinetic energy of a spherical cloud of radius $R$ , density profile  $\rho=\rho(R)\left(\frac{r}{R}\right)^{-\gamma}$, and velocity dispersion profile $\sigma(r)=\sigma(R)\left(\frac{r}{R}\right)^{\beta}$ is given by:

\begin{equation}
E_k(r)=\frac{3}{2}\int_0^r\rho(r)\sigma(r)^24\pi r^2 dr
\end{equation}

\begin{equation}
E_k(r)=6\pi \frac{\rho(R)\sigma(R)^2R^{\gamma-2\beta}}{3+2\beta-\gamma} r^{3+2\beta-\gamma}
\label{Ek}
\end{equation}

\noindent Note that here we chose $r_0=R$ as it will simplify the calculations. In Equation \ref{Ek}, $\rho(R)$ and $\sigma(R)$, the density and velocity dispersion at radius $r=R$  are not observable quantities. But, $\overline{\rho}(R)$ and $\overline{\sigma}(R)$, the average density and mass weighted velocity dispersion within radius $r=R$, are. One therefore needs to derive the relation between $\rho(R)$, $\sigma(R)$ and  $\overline{\rho}(R)$, $\overline{\sigma}(R)$   in order to sub-in the former within the equation of kinetic energy. For the density we have:
\begin{equation}
\overline{\rho}(R)=\frac{\int_0^R \rho(r) 4 \pi r^2dr}{\int_0^R  4 \pi r^2dr}
\end{equation}

\begin{equation}
\overline{\rho}(R)=\rho(R)\frac{3}{3-\gamma}
\end{equation}

\noindent For the velocity dispersion we have:
\begin{equation}
\overline{\sigma}(R)=\frac{\int_0^R \sigma(r)\rho(r) 4 \pi r^2dr}{\int_0^R  \rho(r)4 \pi r^2dr}
\end{equation}

\begin{equation}
\overline{\sigma}(R)=\sigma(R) \frac{3-\gamma}{3+\beta-\gamma}
\end{equation}

\noindent We can now sub these expressions in the equation of kinetic energy for $r=R$:

\begin{equation}
E_{k}(R)=2\pi\overline{\rho}(R)\overline{\sigma}(R)^2R^3\frac{(3+\beta-\gamma)^2}{(3-\gamma)(3+2\beta-\gamma)}
\end{equation}

\begin{equation}
E_{k}(R)=\frac{3}{2} M \overline{\sigma}(R)^2\frac{(3+\beta-\gamma)^2}{(3-\gamma)(3+2\beta-\gamma)}
\end{equation}

The correction factor on the kinetic energy $a_3$ (keeping the same notation as in MD17) resulting from the non-uniform velocity dispersion is therefore given by: 

\begin{equation}
a_3=\frac{(3+\beta-\gamma)^2}{(3-\gamma)(3+2\beta-\gamma)}
\end{equation}

For $\beta=0.5$ and $\gamma=1$, we obtain a correction factor $a_3=\frac{25}{24}$, which is basically negligible. This is quite different to the correction factor evaluated by MD17, which gives $a_3=\frac{2}{3}$ for the same power law indices. After exchanging with the authors, it has been found that MD17 wrongly assumed that the observationally measured velocity dispersion was $\sigma(R)$ as opposed to $\overline{\sigma}(R)$. This mistake leads to the differences in the estimated correction factors highlighted here. Since  $a_3$ is close to unity, we did not take it into account in the virial ratio estimated presented in this article.

 \section{Properties comparison with H09}
 
\begin{figure}
  \vspace{-0.cm}
 \hspace{-.cm}
   \centering
   \includegraphics[width=7.cm]{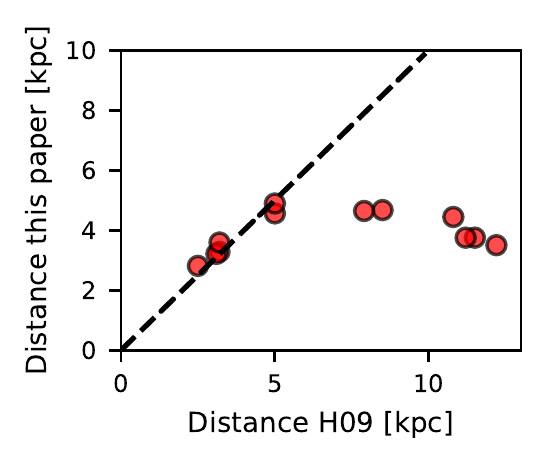} 
\vspace{-0.0cm}
      \caption{Distance comparison for the sub-sample of 12 clouds common to our study, i.e. y-axis, and that of Heyer et al. (2009), i.e. x-axis.}
         \label{distcomp}
   \end{figure}

In Figure \ref{distcomp} we present the comparison of kinematics distances between this paper and Heyer et al. (2009). One can see that for half the clouds we have in common with H09, the distance differ. The reason is that we opted for the near kinematic distances for all of our clouds, while they opted for the far kinematic distance for a significant fraction of their sample. However, as justified in the main body of this paper, it is more likely that all this IRDC-hosting molecular clouds lie at the near distance.  

\begin{figure}
  \vspace{-0.cm}
 \hspace{-.cm}
   \centering
   \includegraphics[width=7.cm]{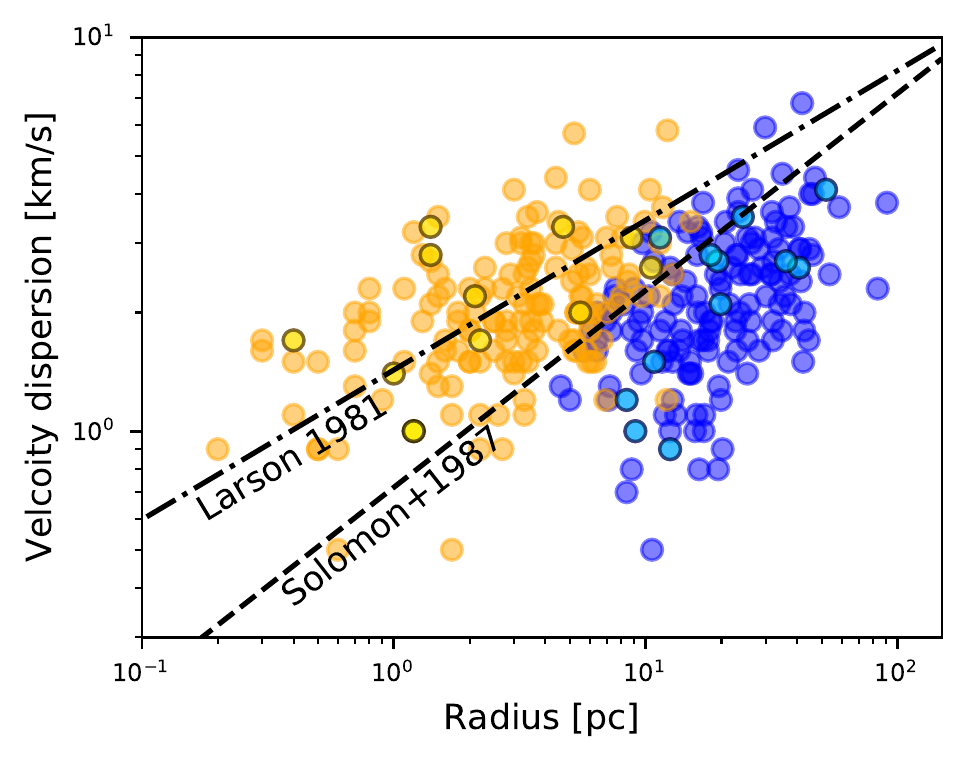} 
\vspace{-0.0cm}
      \caption{Radius versus velocity dispersion for all Heyer al. 2009 measurements. The blue symbols show the measurements done within the original boxes from Solomon et al. 1987. The yellow symbols show the measurements made within the half-power isophot of the column density peak within that box. The 12 clouds that are  in common between our and Heyer's study are shown as yellow and cyan symbols. The dashed and dotted-dashed lines show the Larson's and Solomon's realtions. }
         \label{radvH09}
   \end{figure}

In Figure \ref{radvH09} we present all velocity dispersion measurements as a function of radius from H09. As explained in the main body of the paper, the blue points are cloud-size measurements, while the yellow points are within the FWHM of the emission peak. The points highlighted with solid black circle are those in common between H09 and our sample. The dashed lines show L81 and S87's relationships.

\bsp	
\label{lastpage}
\end{document}